# Intermittent yet coordinated regional strategies can alleviate the COVID-19 epidemic: a network model of the Italian case

Fabio Della Rossa[*,a,b], Davide Salzano[*,b], Anna Di Meglio[*,b], Francesco De Lellis[*,b], Marco Coraggio[b], Carmela Calabrese[b], Agostino Guarino[b], Ricardo Cardona[b], Piero De Lellis[+, b], Davide Liuzza[+,c], Francesco Lo Iudice[+,b], Giovanni Russo[+,d], Mario di Bernardo[b,e,#]

[a]Department of Electronic, Information and Biomedical Engineering, Politecnico di Milano, Italy

[b]Department of Electrical Engineering and Information Technology, University of Naples Federico II, Italy

[c]ENEA, Fusion and Nuclear Safety Department, Frascati (Rome), Italy

[d]Department of Information and Electrical Engineering and Applied Mathematics, University of Salerno, Italy

[e] Department of Engineering Mathematics, University of Bristol, United Kingdom

[*] *These authors contributed equally*

[+] *These authors contributed equally*

[#] Correspondence should be addressed to: mario.di.bernardo@unina.it, m.dibernardo@bristol.ac.uk

Abstract

The COVID-19 epidemic that emerged in Wuhan China at the end of 2019 hit Italy particularly hard, yielding the implementation of strict national lockdown rules (Phase 1). There is now a hot ongoing debate in Italy and abroad on what the best strategy is to restart a country to exit a national lockdown (Phase 2). Previous studies have focused on modelling possible restarting scenarios at the national level, overlooking the fact that Italy, as other nations around the world, is divided in administrative regions who can independently oversee their own share of the Italian National Health Service. In this study, we show that regionalism, and heterogeneity between regions, is essential to understand the spread of the epidemic and, more importantly, to design effective post Lock-Down strategies to control the disease. To achieve this, we model Italy as a network of regions and parameterize the model of each region on real data spanning almost two months from the initial outbreak. Using the model, we confirm the effectiveness at the regional level of the national lockdown strategy implemented so far by the Italian government to mitigate the spread of the disease and show its efficacy at the regional level. We also propose that differentiated, albeit coordinated, regional interventions can be effective in Phase 2 to restart the country and avoid future recurrence of the epidemic, while avoiding saturation of the regional health systems and mitigating impact on costs. Our study and methodology can be easily extended to other levels of granularity (provinces or counties in the same region or states in other federal countries, etc.) to support policy- and decision-makers.

Introduction

Regionalism is an integral part of the Italian constitution. Each of Italy's twenty administrative regions is independent on Health and oversees its own share of the Italian National Health service. The regional presidents and their councils can independently take their own actions, strengthening or, at times, weakening national containment rules. Previous studies have modelled the spread of the epidemics and its evolution in the country at the national level[1-5], and some have looked at the effects of different types of containment and mitigation strategies[6-11]. Limited work[12-21] has taken into account the spatial dynamics of the epidemic but, to the best of our knowledge, no previous paper in the literature has explicitly taken into consideration the pseudo-federalist nature of the Italian Republic and its strong regional heterogeneity when it comes to health matters, hospital capacity, economic costs of a lockdown and the presence of inter-regional people's flows.

In this study, we investigate the whole of the country as a network of regions, each modelled with different parameters. The goal is to identify if and when measures taken by the Italian government had an effect at both the national, but most importantly, at the regional level. Also, we want to uncover the effects on the epidemic spread of regional heterogeneity and inter-regional flows of people and use control theoretic tools to propose and assess differentiated interventions at the regional level to reopen the country and avoid future recurrent epidemic outbreaks.

As aggregate models of the COVID-19 epidemic cannot capture these effects, to carry out our study we derived and parameterized from real data a network model of the epidemics in the country (see **Figure 1a**), where each of the 20 regions is a node and links model both proximity flows and long-distance transportation routes (ferries, train, planes). The model is first shown to possess the right level of granularity and complexity to capture the crucial elements needed to correctly predict and reproduce the available data. Then, it is used to design and test differentiated feedback interventions at the regional level to alleviate the epidemic impact.

Using the model and an *ad hoc* algorithm to parameterize it from real data, we evaluate the effectiveness of the national lockdown strategy implemented so far by the Italian government providing evidence of its efficacy across regions. Also, we show that inter-regional fluxes must be carefully controlled as they can have dramatic effects on recurrent epidemics waves. Finally, we convincingly show that regional feedback interventions, where each of the twenty regions strengthens or weakens local mitigating actions (social distancing, inflow/outflow control) as a function of the saturation of their hospital capacity, can be beneficial in mitigating possible outbreaks and in avoiding recurrent epidemic waves while reducing the costs of a nation-wide lockdown.

Results

Regional effects of the national lock-down

Our approach successfully uncovers the regional effects of the national lock-down measures set in place by the Italian government initially in two northern regions (Lombardy and Veneto from the February 27th 2020), and then nationally from March 8th 2020 till May 4th 2020. We observe that notable parameter changes, detected automatically by our parameterization procedure (see Methods), occur as an effect of such measures with a certain degree of homogeneity across all regions (see **Figures S8, S9** and **Table S4** in the Supplementary Information showing the changes in the social distancing parameter $\rho_i$ over the period of interest). This confirms the effectiveness across the country of the strict social distancing rules implemented at the national level as also noted in previous work[1,2,14] modeling the country as a whole.

The representative examples of two regions, Lombardy in the North and Campania in the South, highlighted in **Figure 1,** show that the model is able to correctly capture the effect of such measures in both the regions, see **Table 1**. The model also captures the effect of the flow of people that travelled from North to South when the national lockdown measures where first announced on March 8th 2020. As shown in **Table 1**, the estimated number of infected predicted by the model for the Campania region in the time window March 19th - March 30th 2020 is detected to suddenly increase at the beginning of the next time window. This can be explained as a possible effect of the movement of people from North to South that occurred around 15 days before. Also, the data analysis consistently shows across many regions that the mortality rate changes dramatically as a function of the saturation of the hospital beds in the region (see **Figure S11** and **Section S2** of the Supplementary Information for further details).

Regional heterogeneity counts

After confirming the predictive and descriptive ability of the proposed model, we investigated next the influence of regional heterogeneity on the onset of an epidemic outbreak and the occurrence of possible recurrent epidemic waves. To this aim we set the model with parameters capturing the situation in each region on May 3$^{rd}$ 2020, when the effects of the national lock down were fully in place, and simulated the scenario where just one of the twenty regions (e.g., Lombardy in the North of Italy) fully relaxes its lockdown. As reported in **Figure 2**, we found that a primary outbreak in that region would quickly propagate causing secondary recurrent outbreaks in other regions including Emilia-Romagna and Piedmont. At the national level this would cause the onset of a second epidemic wave that, if not contained, would end up afflicting more than 25% of the entire population. An even more dramatic scenario would emerge if inter-regional flows where concurrently restored to their pre-lockdown levels (see **Figure S1**) or all regions were to relax their current restrictions concurrently (see **Figure S2** in the Supplementary information).

**Feedback regional interventions can be beneficial**

A crucial open problem is to support decision makers in determining what form of interventions might be beneficial to avoid the onset of future outbreaks while mitigating the cost of Draconian interventions at the national level. To this aim we compared the effects of national measures (e.g., general lockdown) against those of a regional feedback strategy where social distancing measures are put in place or relaxed independently by each region according to the ratio between hospitalized individuals and the total capacity of the health system in that region. In particular, we assume each region implements a stricter lockdown when such a ratio becomes greater or equal than 20% and relaxes the social distancing rules when it is below 5% (see **Methods** for further details). **Figure 3** confirms the effectiveness of such a local strategy, where we see that a differentiated strategy among the regions is more effective than a national lockdown in avoiding future waves of the disease (**Figure S3**) but, most importantly, also in guaranteeing that no region exceeds its own hospitals' capacity. Moreover, intermittent regional measures yield lower economic costs for the country, as regional economies can be restarted and remain open for a much longer time (**Table 2** and **Table S1**); an effect that becomes even more apparent when regions concurrently increase their testing capacity as shown in **Figure 4** and reported in **Table 2.**

Discussion

Following the initial COVID-19 outbreak in Northern Italy, the Italian government, as many other governments around the world, adopted increasingly stricter lockdown measures at the national level to mitigate the epidemic. Despite their success, their tolling economic costs have stirred a hot national debate on whether such measures were necessary in the first place and on how to relax them while avoiding future epidemic waves. Several attempts have been made in the literature at addressing this pressing open issues by means of aggregate models (originated from the classical SIR model) to describe the effects of different intervention strategies at the national level[6,7]. A network model has also been recently proposed to describe the spatial dynamics of the spread of the COVID-19 epidemics among the 107 Italian provinces[14]. Other works in the literature have explored the effects of intermittent measures, either periodic or as a function of some observable quantities, as a viable alternative to long, continuous periods of national lockdown. However, the effects of these strategies have only been investigated on theoretical aggregate models at the national level[7,22].

An important missing aspect that we considered in our study is the effect of regional heterogeneity on the efficacy of the measures taken so far and the possibility of adopting differentiated and localized intervention strategies thanks to the pseudo-federalist administrative structure of the Italian Republic. Our results confirm the effectiveness at the regional level of the national lockdown measures taken so far. They also convincingly reveal the presence of important regional effects due, for example, to the saturation of regional healthcare systems or to the presence of notable North-South flows in the country that followed the announcement of national measures. Also, contrary to previous work, we explicitly accounted for the strongly nonlinear nature of the model and the uncertainty present in the data by performing a sensitivity analysis on the estimated parameters that further confirmed the robustness of the proposed strategies for a wide range of parameter changes

Our study strongly suggests for policy- and decision- makers the potential benefits of differentiated (but coordinated) feedback regional interventions, which can be used independently or in combination with other measures, in order to avoid future epidemic waves or even to contain the outbreak of potential future epidemics. Despite having been focused on Italy, our methodology and modeling approach can be easily extended to other levels of granularity, e.g. countries in a continent or counties in a state, and adapted to any other nation where regional heterogeneity is important and cannot be neglected; notable examples are countries with a federal state organization such as Germany or the United States of America.

Future work needs to address further aspects as for example exploring how the structural properties of the inter-regional network can influence the dynamics of the epidemic or adopting more sophisticated cost functions to design more effective region-specific mitigation strategies in other contexts or for other purposes.

# Methods

## Regional and national model

As a *regional model* of the COVID-19 epidemic spread we use the compartmental model shown in **Figure 4** which we found from data analysis and identification trials to be the simplest model structure able to capture the real data. The full model equations describe the dynamics of susceptible ($S_i$), infected ($I_i$), quarantined ($Q_i$), hospitalized ($H_i$), recovered ($R_i$) and deceased ($D_i$) and are given as:

$$\dot{S}_i = -\rho_i \beta \frac{S_i I_i}{N_i}, \tag{1}$$

$$\dot{I}_i = \rho_i \beta \frac{S_i I_i}{N_i} - \alpha_i I_i - \psi_i I_i - \gamma I_i, \tag{2}$$

$$\dot{Q}_i = \alpha_i I_i - \kappa_i^H Q_i - \eta_i^Q Q_i + \kappa_i^Q H_i \tag{3}$$

$$\dot{H}_i = \kappa_i^H Q_i + \psi_i I_i - \eta_i^H H_i - \zeta_i H_i - \kappa_i^Q H_i \tag{4}$$

$$\dot{D}_i = \zeta_i H_i, \tag{5}$$

$$\dot{R}_i = \gamma I_i + \eta_i^Q Q_i + \eta_i^H H_i \tag{6}$$

where $\beta$ is the infection rate which will be assumed to be the same for all regions since COVID-19 is transmitted from person to person and there is no parasite vector or evidence of environmental parameters significantly altering its infection rate, $\rho_i \in [0,1]$ is a parameter modeling the effects of social distancing measures in the $i$-th region, $\alpha_i$ is the rate of infected that are detected and quarantined, $\psi_i$ is the rate of infected that needs to be hospitalized, $\gamma$ is the recovery rate of the infected that is assumed to be equal for all regions, $\eta_i^Q$ is the rate of quarantined who recover, $\eta_i^H$ is the fraction of hospitalized who recover, $\kappa_i^Q$ is the rate of hospitalized that is transferred to home isolation, $\kappa_i^H$ is the rate of quarantined who need to be hospitalized, and $\zeta_i$ is the mortality rate that was shown from data analysis (see Section S3 of the Supplementary Information) to be a function of the ratio between $H_i$ and the maximum number, say $T_i^H$, of patients that can be treated in ICU at the hospitals in $i$-th region. $N_i$ is the actual population in the $i$-th region, i.e. the resident population without those removed because quarantined, hospitalized, deceased or recovered.

Extending previous approaches for modelling Dengue fever in Brazil[23], we obtain the *national network model* of the COVID-19 epidemic in Italy as a network of twenty regions (see **Figure 1a**) interconnected by links modeling commuter flows and major transportation routes among them.

The network model of Italy we adopt in this study is, for $i = 1, \dots, 20$,

$$\dot{S}_i = -\sum_{j=1}^{M}\sum_{k=1}^{M} \rho_j \beta \phi_{ij}(t) S_i \frac{\phi_{kj}(t) I_k}{N_j^p}, \tag{7}$$

$$\dot{I}_i = \sum_{j=1}^{M}\sum_{k=1}^{M} \rho_j \beta \phi_{ij}(t) S_i \frac{\phi_{kj}(t) I_k}{N_j^p} - \alpha_i I_i - \psi_i I_i - \gamma I_i, \tag{8}$$

$$\dot{Q}_i = \alpha_i I_i - \kappa_i^H Q_i - \eta_i^Q Q_i + \kappa_i^Q Q_i, \tag{9}$$

$$\dot{H}_i = \kappa_i^H Q_i + \psi_i I_i - \eta_i^H H_i - \kappa_i^Q H_i - \zeta\left(H_i/T_i^H\right) H_i, \tag{10}$$

$$\dot{D}_i = \zeta\left(H_i/T_i^H\right) H_i, \tag{11}$$

$$\dot{R}_i = \gamma I_i + \eta_i^Q Q_i + \eta_i^H H_i \tag{12}$$

$$N_i^p = \sum_{k=1}^{M} \phi_{ki}(t)(S_k + I_k + R_k) \tag{13}$$

where in addition to the parameters and states described above, we included the fluxes $\phi_{ij}(t)$ between regions; $\phi_{ij}(t): \mathbb{R} \to [0,1]$ denoting the ratio of people from region $i$ interacting with those in region $j$ at time $t$, such that $\sum_j \phi_{ij}(t) = 1$.

### Model parameterization from real data and model validation

We divide the model parameterization into two stages. Firstly, we estimate from the available data the parameters of each of the twenty regional models; then, we use publicly available mobility data in Italy to estimate the fluxes among the regions.

We follow the current consensus in the literature on COVID-19[2] (see **Table S5**) by setting $\beta = 0.4$ and $\gamma = 1/14$ for all regions. We make the ansatz that parameters remain constant over time intervals $T_k$ but do not assume the duration of such intervals known *a priori*. Therefore, we set the problem of estimating the parameters values and *when* they change in each region (as a likely result of national containment measures). We start estimating parameters in a region from the first date when the number of deceased and the number of recovered is greater or equal than 10.

Noting that the available data for the COVID-19 epidemics, as collected by the Dipartimento della Protezione Civile - Presidenza del Consiglio dei Ministri (the Italian Civil Protection Agency), includes for each region the daily numbers of quarantined ($\tilde{Q}_i$), hospitalized ($\tilde{H}_i$), deceased ($\tilde{D}_i$) and the recovered from those who were previously hospitalized or quarantined, say $\tilde{R}_i^O$, we discretize and rewrite model (1)-(6) for each region ($i$ = 1,…,20) as (dropping the subscripts to the parameters for notational convenience)

$$\hat{S}_i(t+1) = \hat{S}_i(t) - \rho\beta \frac{\hat{S}_i(t)\hat{I}_i(t)}{N_i(0) - \tilde{Q}_i(t) - \tilde{H}_i(t) - \tilde{D}_i(t)} \tag{14}$$

$$\hat{I}_i(t+1) = \hat{I}_i(t) + \rho\beta \frac{\hat{S}_i(t)\hat{I}_i(t)}{N_i(0) - \tilde{Q}_i(t) - \tilde{H}_i(t) - \tilde{D}_i(t)} - \gamma\hat{I}_i(t) - \tau\hat{I}_i(t) \tag{15}$$

$$\hat{C}_i(t+1) = \tilde{C}_i(t) + \tau\hat{I}_i(t) \tag{16}$$

$$\hat{Q}_i(t+1) = \tilde{Q}_i(t) + \alpha\hat{I}_i(t) - \eta^Q \tilde{Q}_i(t) - \kappa^H \tilde{Q}_i(t) + \kappa^Q \tilde{H}_i(t) \tag{17}$$

$$\hat{H}_i(t+1) = \tilde{H}_i(t) + \psi\hat{I}_i(t) - \eta^H \tilde{H}_i(t) + \kappa^H \tilde{Q}_i(t) - \kappa^Q \tilde{H}_i(t) - \zeta\tilde{H}_i(t) \tag{18}$$

$$\hat{R}_i^O(t+1) = \tilde{R}_i^O(t) + \eta^Q \tilde{Q}_i(t) + \eta^H \tilde{H}_i(t) \tag{19}$$

$$\hat{D}_i(t+1) = \tilde{D}_i(t) + \zeta\tilde{H}_i(t) \tag{20}$$

where measured quantities are denoted by a tilde and estimated state variables by a hat and $\tau := \alpha + \psi$. Here, $C_i = Q_i + H_i + D_i + R_i^O$ represents the total number of cases detected in region *i* as daily announced by the Protezione Civile.

We notice that, exploiting the available data, the predictor can be split into two parts so that two different algorithms can then be used to estimate the parameters of each part. An *ad hoc* identification algorithm estimates the parameters of the nonlinear part (Equations (14)-(16)) and automatically detects the breakpoints where notable parameter changes occur, while an ordinary least squares method is then used to identify the parameters of for the linear one (Equations (17)-(20)), as described in detail in **Section S2** of the Supplementary Information. Note that, as the actual number of infected is not known[9,24], we include the number of infected at the beginning of each time window as a parameter to be estimated by the algorithm used for the nonlinear part.

The results of the identification process also show the presence of a statistically significant correlation (*p*-value equal to 0.034) between the value of the mortality rate, parameter $\zeta_i$ in model (1)-(6), and the saturation of the regional health system represented by the ratio between the number of hospitalized in that region ($H_i$) and the total number of available hospital beds ($T_i^H$) (See Supplementary Information **Section S2** and **Figure S11** for further details and function estimation).

Validation is carried out by using the parameterized model to capture the available data for each window showing an average quadratic error less than 10% over the entire dataset. The parameters identified in each window can also be used to provide model predictions of future trends of the epidemic diseases as discussed in **Section S2** of the Supplementary Information.

Cost Estimation

We estimate the cost of each regional lockdown as the sum of the costs for social care and the loss of added value. The costs for social care in each region were computed as the costs for layoff support (``cassa integrazione in deroga''), estimated by multiplying the number of requests[26] by 65% of the average regional monthly income[27], together with the non-repayable-loan of 600 € given to self-employed workers by the Italian Government during the national lockdown[28]. The loss of added value per day was taken from the values estimated by SVIMEZ (the Italian Association for the development of Industry in the South) in **Table 3** of their online report[29]. We then compute the daily costs of the lockdown and use it to estimate the total costs of each of the simulated scenarios.

Data fitting and sensitivity analysis

All computational analyses and the fitting of data were performed using MATLAB and its optimization toolbox. To account for the inherent uncertainty associated to the COVID-19 epidemic, each result reported in the manuscript is the output of 10,000 numerical simulations, where we varied the values of the model parameters using the Latin Hypercube sampling method[25]. Specifically, the regional parameters $\alpha_i$, $\psi_i$, $\kappa_i^Q$, $\kappa_i^H$, $\eta_i^Q$, $\eta_i^H$ and the estimated initial condition at May 3rd 2020 $I_{f,i}$ were varied considering a maximum variation of $\pm 20\%$ from their nominal values (indicated in **Table S4** of the Supplementary Information).

Implementation and design of national and regional feedback intervention strategies

We model the implementation of *regional social distancing strategies* by capturing their effects as a variation of the social distancing parameters, $\rho_i$ in (7)-(8), in each region. Specifically, we assume each region follows the feedback control rule:

$$\rho_i = \begin{cases} \underline{\rho_i}, & \text{if } \dfrac{0.1 H_i}{T_i^H} \geq 0.20 \\ \bar{\rho}_i, & \text{if } \dfrac{0.1 H_i}{T_i^H} \leq 0.05 \end{cases}$$

where $\underline{\rho_i}$ is set equal to the minimum estimated value in that region during the national lockdown (see Table S9) and $\bar{\rho}_i$ increased as a worst case to $\min(1, 3\underline{\rho_i})$ so as to simulate the effect of relaxing the lock-down measures in each region. (The case where $\bar{\rho}_i$ is set to a lower value equal to $1.5\underline{\rho_i}$ is shown for the sake of comparison in **Figure S5** and **Figure S6** of the Supplementary Information).

Also, when a region is shut down, we assume all fluxes in and out of that region are reduced to 70% of their original values to better simulate the actual reduction in people's movement observed during the lockdown in Italy (for further details see SI).

National lockdown measures are modelled by setting all $\rho_i$ simultaneously to $\underline{\rho_i}$ in all regions and reducing all fluxes by 70% while national reopening of all regions by setting all $\rho_i$ simultaneously to $\bar{\rho}_i$ and restoring interregional flows to their pre-lockdown level.

To model the increase in the COVID-19 testing capacity of each region the parameter $\alpha_i$ in region $i$ is multiplied by a factor 2.5 which corresponds to the average increase in the number of tests carried out nationally since the COVID-19 outbreak first started.

### Numerical simulations

All simulations were carried out in MATLAB with a discretization step of 1 day to match the available data sampling interval. Initial conditions for regional compartments were set as follows. Quarantined, Hospitalized, Deceased and Recovered are initially set to the datapoints available for May 3rd 2020. The number of infected is set to the value $I_f$ estimated by our procedure for that date while Susceptibles are initialized to the resident population form which the other compartments are removed.

The code is available at

https://github.com/diBernardoGroup/Network-model-of-the-COVID-19

Further details are given in Section S5 of the Supplementary Information.

## Data availability

## Acknowledgements

MdB, M.C. and A.G. wish to acknowledge support from the European Project FET-OPEN Cosy-Bio. MdB, M.C., P.D.L. and F.L.I acknowledge support from the research project PRIN 2017 "Advanced Network Control of Future Smart Grids" funded by the Italian Ministry of University and Research (2020-2023). P.D.L. and F.D.R. acknowledge support from the University of Naples Federico II and the Compagnia di San Paolo, Istituto Banco di Napoli - Fondazione for supporting their research under the grant STAR 2018, project ACROSS.

## Author Contributions

M.d.B with support from P.D.L., D.L., F.D.R., G.R. and F.L.I. designed the research; F.D.R., A.D.M, P.D.L. and D.S. carried out the model parameterization; F.L.I., C.C. and A.G. analyzed the data and estimated the inter-regional flux matrix; F.D.L., M.C. and R.C. wrote and checked the numerical code used for all simulations; D.L., M.C. and R.C. investigated feedback strategies for mitigation and containment; F.D.L., D.S., M.C., C.C., A.G. and R.C. carried out the numerical simulations with inputs from the rest of the authors; M.d.B. wrote the manuscript with inputs from F.D.R., P.D.L., A.D.M., D.L., F.L.I. and G.R.

## Competing Interests

The authors declare no competing interests.

# References

[1] Calafiore, G. C., Novara, C., & Possieri, C. A Modified SIR Model for the COVID-19 Contagion in Italy. Preprint at https://arxiv.org/abs/2003.14391 (2020).

[2] Giordano, G., Blanchini, F., Bruno, R., Colaneri, P., Di Filippo, A., Di Matteo, A., & Colaneri, M. Modelling the COVID-19 epidemic and implementation of population-wide interventions in Italy. *Nature Medicine*, 1–6 (2020).

[3] Hethcote, H. W. The Mathematics of Infectious Diseases. *SIAM Review* **42**, 599–653 (2000).

[4] Mummert, A., & Otunuga, O. M. Parameter Identification for a Stochastic SEIRS Epidemic Model: Case Study Influenza. *Journal of Mathematical Biology* **79**, 705–729 (2019).

[5] Peng, L., Yang, W., Zhang, D., Zhuge, C., & Hong, L. Epidemic Analysis of COVID-19 in China by Dynamical Modeling. Preprint at https://arxiv.org/abs/2002.06563 (2020).

[6] Casella, F. Can the COVID-19 Epidemic Be Controlled on the Basis of Daily Test Reports? Preprint at https://arxiv.org/abs/2003.06967 (2020).

[7] Bin, M., Cheung, P., Crisostomi, E., Ferraro, P., Lhachemi, H., Murray-Smith, R., Myant, C., Parisini, T., Shorten, R., Stein, S. & Stone, L. On Fast Multi-Shot Epidemic Interventions for Post Lock-Down Mitigation: Implications for Simple Covid-19 Models. Preprint at https://arxiv.org/abs/2003.09930 (2020).

[8] Imperial College COVID-19 Response Team, Report 9: Impact of Non-Pharmaceutical Interventions (NPIs) to Reduce COVID-19 Mortality and Healthcare Demand. Available at https://www.imperial.ac.uk/mrc-global-infectious-disease-analysis/covid-19/report-9-impact-of-npis-on-covid-19/ (2020)

[9] Imperial College COVID-19 Response Team, Report 13: Estimating the Number of Infections and the Impact of Non-Pharmaceutical Interventions on COVID-19 in 11 European Countries. Available at https://spiral.imperial.ac.uk/handle/10044/1/77731 (2020).

[10] Nowzari, C., Preciado, V. M., & Pappas, G. J. Optimal Resource Allocation for Control of Networked Epidemic Models. *IEEE Transactions on Control of Network Systems* **4**, 159–169 (2015).

[11] Ramírez-Llanos, E., & Martínez, S. A Distributed Dynamics for Virus-Spread Control. *Automatica* **76**, 41–48 (2017).

[12] Arino, J., & Van den Driessche, P. A Multi-City Epidemic Model. *Mathematical Population Studies* **10**, 175–193 (2003).

[13] Ganesh, A., Massoulie, L., & Towsley, D. The Effect of Network Topology on the Spread of Epidemics. *Proceedings IEEE 24th Annual Joint Conference of the IEEE Computer and Communications Societies* **2**, 1455-1466 (2005).

[14] Gatto, M., Bertuzzo, E., Mari, L., Miccoli, S., Carraro, L., Casagrandi, R., & Rinaldo, A. Spread and Dynamics of the COVID-19 Epidemic in Italy: Effects of Emergency Containment Measures. *Proceedings of the National Academy of Sciences* (2020), https://doi.org/10.1073/pnas.2004978117

[15] Hethcote, H. W. An Immunization Model for a Heterogeneous Population. *Theoretical Population Biology* **14**, 338–349 (1978).

[16] Lajmanovich, A., & Yorke, J. A. A Deterministic Model for Gonorrhea in a Nonhomogeneous Population. *Mathematical Biosciences* **28**, 221–236 (1976).

[17] Mei, W., Mohagheghi, S., Zampieri, S., & Bullo, F. On the Dynamics of Deterministic Epidemic Propagation over Networks. *Annual Reviews in Control* **44**, 116–128 (2017).

[18] Sattenspiel, L. & Dietz, K. A Structured Epidemic Model Incorporating Geographic Mobility among Regions. *Mathematical Biosciences* **128**, 71–92 (1995).

[19] Scarpino, S. V., & Petri, G. On the Predictability of Infectious Disease Outbreaks. *Nature Communications* **10**, 898 (2019).

[20] Van Mieghem, P., Omic, J., & Kooij, R. Virus Spread in Networks. *IEEE/ACM Transactions on Networking* **17**, 1–14 (2008).

[21] Wang, W., & Zhao, X.-Q. An Epidemic Model in a Patchy Environment. *Mathematical Biosciences* **190**, 97–112 (2004).

[22] Meidan, D., Cohen, R., Haber, S. and Barzel, B., 2020. An Alternating Lock-down Strategy for Sustainable Mitigation of COVID-19. *arXiv preprint* arXiv:2004.01453.

[23] Stolerman, L. M., Coombs, D., & Boatto, S. SIR-Network Model and its Application to Dengue Fever. *SIAM Journal on Applied Mathematics* **75**, 2581–2609 (2015).

[24] Li, R., Pei, S., Chen, B., Song, Y., Zhang, T., Yang, W., & Shaman, J. Substantial Undocumented Infection Facilitates the Rapid Dissemination of Novel Coronavirus (SARS-CoV2). *Science*, eabb3221 (2020).

[25] Helton, J. C., & Davis, F. J. (2003). Latin Hypercube Sampling and the Propagation of Uncertainty in Analyses of Complex Systems. Reliability Engineering & System Safety, 81(1), 23-69.

[26] INPS (2020). Dati al 10 maggio su Cassa integrazione ordinaria, assegno ordinario, richieste di pagamento SR41 e Cassa integrazione in deroga https://www.inps.it/nuovoportaleinps/default.aspx?itemdir=53641

[27] Today economia (2018). Stipendi, l'Italia è uno 'scivolo': Regioni e province dove si guadagna più, http://www.today.it/economia/stipendi-italia-2018-regioni-province.html

[28] INPS (2020). Osservatori statistici e altre statistiche, Indennità 600 euro, https://www.inps.it/docallegatiNP/Mig/Dati_analisi_bilanci/StatInBreve_indennita600.pdf

[29] SVIMEZ Associazione per lo sviluppo dell'industria nel mezzogiorno, L'impatto economico e sociale del Covid-19: Mezzogiorno e Centro-Nord, 2020, http://lnx.svimez.info/svimez/wp-content/uploads/2020/04/svimez_impatto_coronavirus_bis.pdf

FIGURES AND CAPTIONS

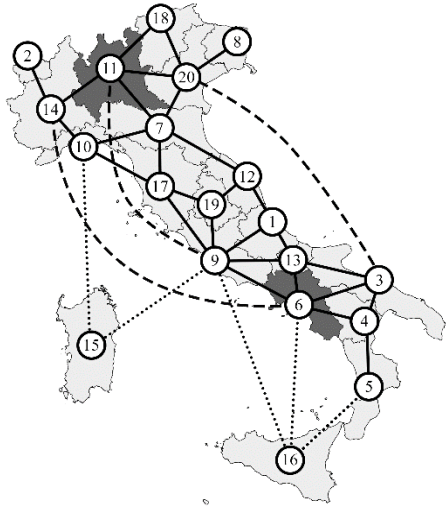

a

b

| Index | Region | Index | Region |
|---|---|---|---|
| 1 | Abruzzo | 11 | Lombardy |
| 2 | Aosta Valley | 12 | Marche |
| 3 | Apulia | 13 | Molise |
| 4 | Basilicata | 14 | Piedmont |
| 5 | Calabria | 15 | Sardinia |
| 6 | Campania | 16 | Sicily |
| 7 | Emilia | 17 | Tuscany |
| 8 | Friuli-Venezia Giulia | 18 | Trentino-Alto Adige |
| 9 | Lazio | 19 | Umbria |
| 10 | Liguria | 20 | Veneto |

**Figure 1. Schematic diagram of the network model structure and representative regional parameters. a.** Representative graph of the network model structure used in the paper. Only a subset of all links is shown for the sake of clarity (the complete graphs are depicted in **Figure S8** of the Supplementary Information). Solid lines represent proximity links, dashed lines long distance transportation routes (air, train, road), dotted lines show major ferry routes between insular regions and the Italian mainland. **b.** Table of the Italian region names and their positions in the graph.

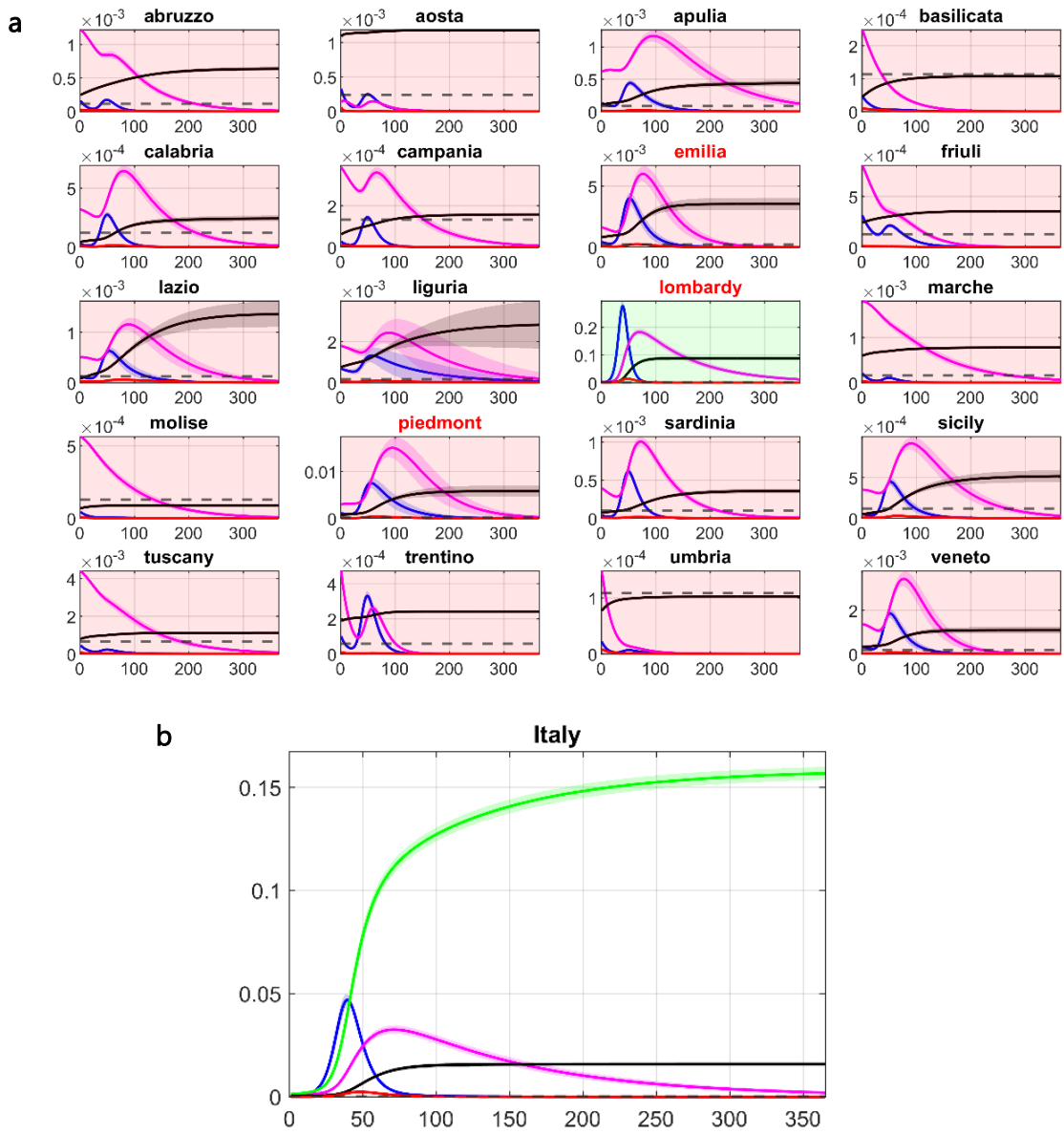

Figure 2. [Phase 2, only one region relaxes its lockdown] a. Regional and b. national dynamics in the case where only one region (Lombardy in North Italy) relaxes its containment measures at time 0, while interregional fluxes are set to the level they reached during the lockdown level. (Regional dynamics when the fluxes between regions are set to their pre-lockdown level are shown in **Figure S1** of the Supplementary Information showing even more dramatic scenarios). Panels of regions adopting a lockdown are shaded in red while those of regions relaxing social containment measures are shaded in greed. Blue, magenta, red, green, and black solid lines correspond to the fraction in the population of infected, quarantined, hospitalized, recovered, and deceased over the entire regional (panels **a**) or national (panel **b**) population averaged over 10,000 simulations with parameters sampled using a Latin Hypercube technique (see Methods) around their nominal values set as those estimated in the last time window for each region as reported in **Table S4** of the Supplementary Information. Shaded bands correspond to twice the standard deviation. The black dashed line identifies the fraction of the population that can be treated in ICU ($T_i^H/N_i$). The regions identified with a red label are those where the total hospital capacity is saturated.

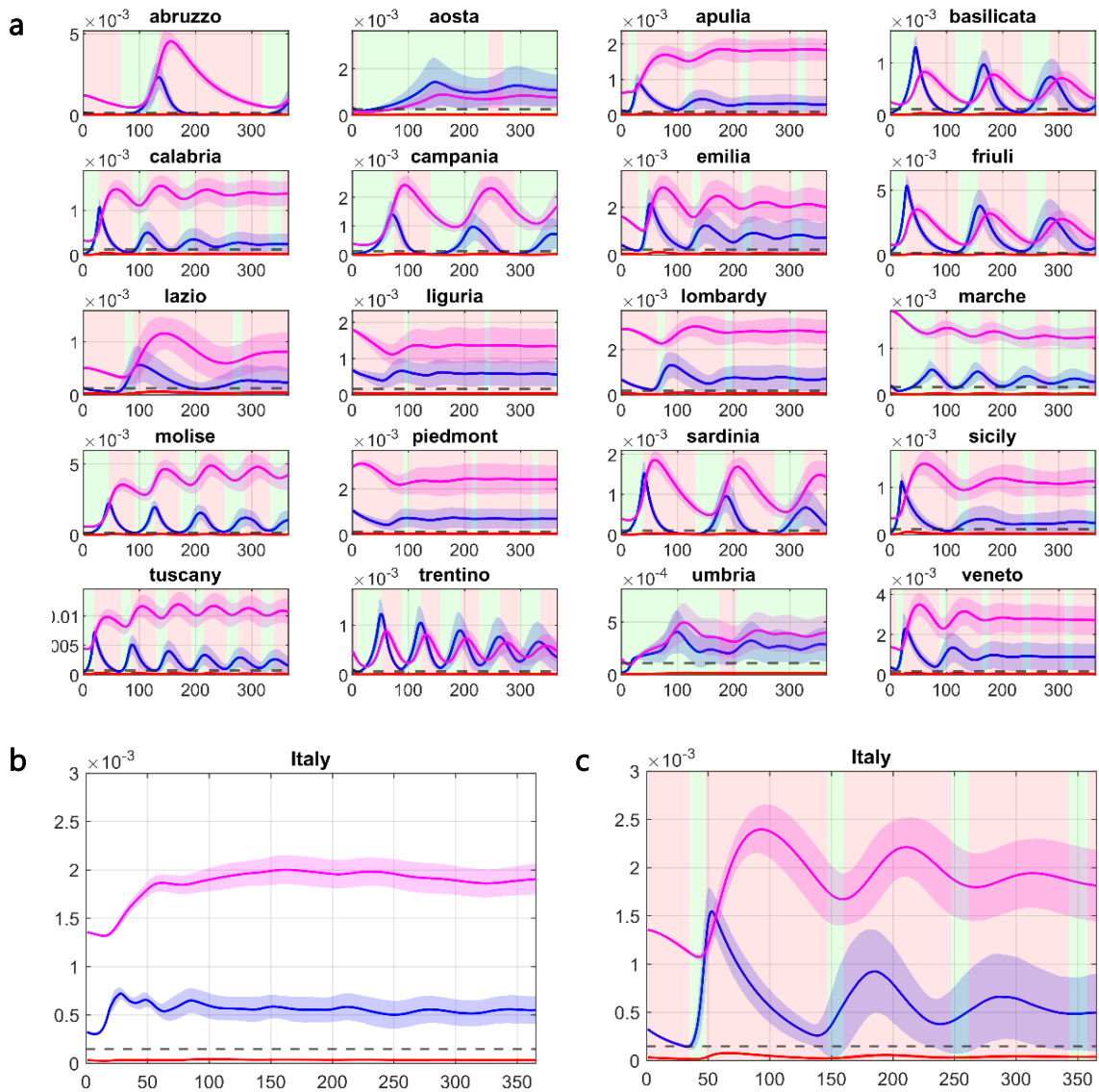

Figure 3. [Phase 2, intermittent regional measures] **a.** Each of the 20 panels shows the evolution in each region of the fraction in the population of infected (blue), quarantined (magenta) and hospitalized requiring ICUs (red) averaged over 10,000 simulations with parameters sampled using a Latin Hypercube technique (see **Methods**) around their nominal values set as those estimated in the last time window for each region as reported in **Table S4** of the Supplementary Information. Shaded bands correspond to twice the standard deviation. Dashed black lines represent line the fraction of the population that can be treated in ICU ($T_i^H/N_i$). Regions adopt lockdown measures in the time windows shaded in red while relax them in those shaded in green. During a regional lockdown, fluxes in/out of the region are set to their minimum level. **b.** National evolution of the fraction in the population of infected (blue), quarantined (magenta) and hospitalized requiring ICUs (red) obtained by summing those in each of the 20 regions adopting intermittent regional measures. **c.** National evolution of the fraction in the population of infected (blue), quarantined (magenta) and hospitalized requiring ICUs (red) when an intermittent national lock down is enforced with all regions shutting down when the total number of occupied ICU beds at the national level exceed 20%. Regional dynamics corresponding to this scenario are shown in **Figure S4** of the Supplementary Information.

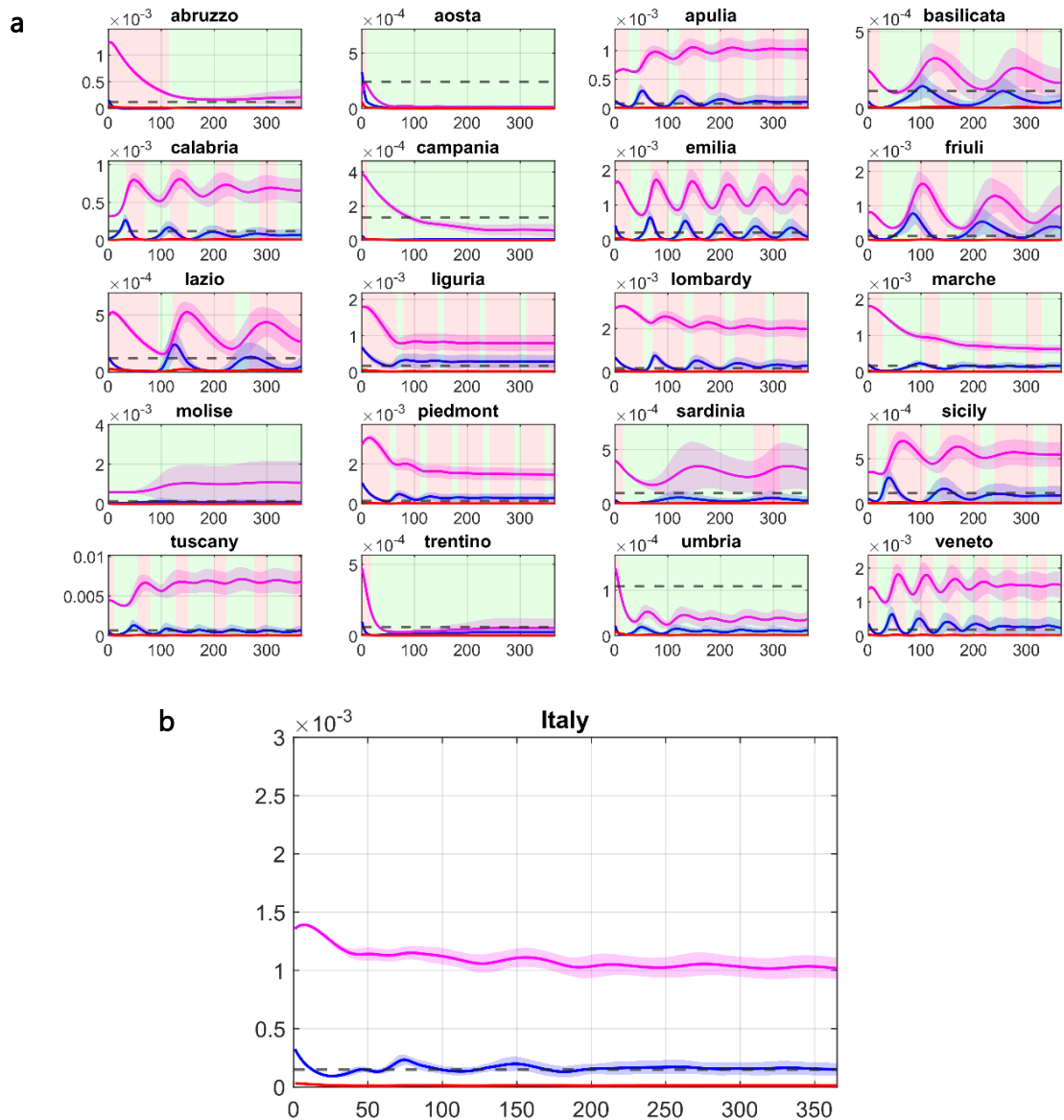

Figure 4. [Phase 2, intermittent regional measures with increased COVID-19 testing capacity] a. Each of the 20 panels shows the evolution in the region named above the panel of the fraction in the population of infected (blue), quarantined (magenta) and hospitalized requiring ICUs (red) averaged over 10,000 simulations with parameters sampled using a Latin Hypercube technique (see **Methods**) around their nominal values set as those estimated in the last time window for each region as reported in **Table S4** of the Supplementary Information. Shaded bands correspond to twice the standard deviation. Dashed black lines represent the fraction of the population that can be treated in ICU ($T_i^H/N_i$). Regions adopt lockdown measures in the time windows shaded in red while relax them in those shaded in green. During a regional lockdown, fluxes in/out of the region are set to their minimum level. Regions COVID-19 testing capacities are assumed to be increased by a factor 2.5 (see **Methods**) with respect to their current values. b. National evolution of the fraction of infected (blue), quarantined (magenta) and hospitalized requiring ICUs (red) obtained by summing those in each of the 20 regions adopting intermittent regional measures

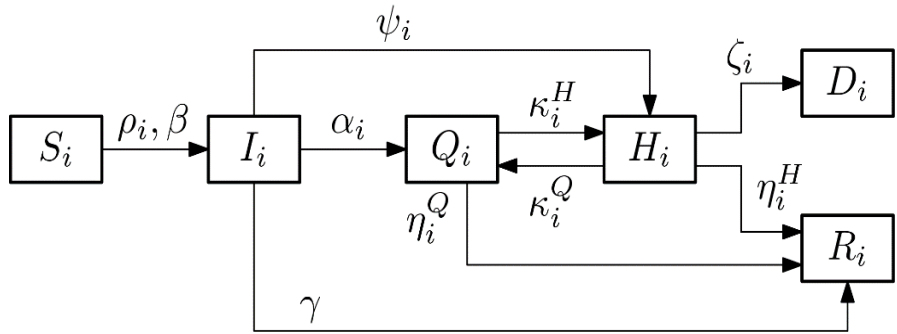

Figure 5. [Regional compartmental model structure adopted in our study]. Schematic structure of model described by equations (1)-(6). Compartments describe the dynamics of susceptible ($S_i$), infected ($I_i$), quarantined ($Q_i$), hospitalized ($H_i$), recovered ($R_i$) and deceased ($D_i$). The number of hospitalized requiring ICU is estimated as 10% of the total. The model structure was suggested from data analysis with links between compartments being added or removed according to how the data was matched by the model (see **Section S3** of the Supplementary Information for further details).

**TABLES**

| Region | Breakpoint | $\rho_i$ | $\alpha_i$ | $\psi_i$ | $\kappa_i^H$ | $\kappa_i^Q$ | $\eta_i^Q$ | $\eta_i^H$ | $\zeta_i$ | $I_0$ | $I_f$ | $R_{0,i}$ |
|---|---|---|---|---|---|---|---|---|---|---|---|---|
| Campania | 3/19/20 | 0.467 | 0.014 | 0.064 | 0.000 | 0.100 | 0.018 | 0.000 | 0.022 | 1231 | 1816 | 1.26 |
|  | 3/30/20 | 0.221 | 0.067 | 0.019 | 0.006 | 0.040 | 0.018 | 0.000 | 0.011 | 2231 | 234 | 0.57 |
| Lombardy | 2/27/20 | 0.742 | 0.008 | 0.073 | 0.000 | 0.052 | 0.022 | 0.044 | 0.033 | 1679 | 30797 | 1.69 |
|  | 3/19/20 | 0.308 | 0.017 | 0.059 | 0.000 | 0.048 | 0.022 | 0.017 | 0.023 | 28715 | 5774 | 0.84 |

**Table 1. Estimated parameter values** for Campania in the South (region no. 6) and Lombardy in the North (region no. 11), where the initial outbreak occurred. These regions are highlighted in a darker colour in **Figure 1**. $I_0$ and $I_f$ are the number of infected estimated in the region at the beginning and at the end of each time window. The first breakpoint is the date when 10 deaths and 10 recovered were first reported in the region and the analysis started. The second breakpoint is the end of the first window and the start of the second window (ending on May 3rd 2020).

| Simulation | Total cases | Total deaths | Maximum hospitalized | Days over hospital's capacity (nation) | Regions over hospital's capacity | Economic cost [M€] |
|---|---|---|---|---|---|---|
| All regions but Lombardy are locked down (Fig. 2) | 10,545,380 ± 146,458 | 956,442 ± 68,215 | 144,180 ± 10,100 | 78.6 ± 2.3 | 3 | 475,330 ± 0 |
| Intermittent regional measures (Fig. 3a,b) | 2,165,229 ± 83,806 | 154,878 ± 3,008 | 2,927 ± 183 | 0 ± 0 | 0 | 470,735 ± 6,353 |
| Intermittent national measure (Fig. 3c) | 2.197.076 ± 189.948 | 176.210 ± 8.264 | 4,794 ± 309 | 0 ± 0 | 3 | 532,802 ± 12,474 |
| Intermittent regional measures with increased testing (Fig. 4) | 939,091 ± 39,414 | 74,017 ± 1,325 | 1,915 ± 0 | 0 ± 0 | 0 | 349.963 ± 9,732 |

Table 2. Comparison of each of the simulated scenarios. Metrics to evaluate the impact over 1 year of each of the simulated scenarios are reported showing the effectiveness of the intermittent regional measures shown in **Figure 3** and **Figure 4** in avoiding any saturation of the regional health systems while mitigating the impact of the epidemic. Average values are shown ±1 standard deviation calculated from 10,000 repetitions of each simulation using parameter values sampled using a Latin Hypercube technique around the nominal parameter values reported in **Table S4**.

# SUPPLEMENTARY INFORMATION

*Intermittent yet coordinated regional strategies can alleviate the COVID-19 epidemic: a network model of the Italian case*

Fabio Della Rossa[*,a], Davide Salzano[*,b], Anna Di Meglio[*,b], Francesco De Lellis[*,b], Marco Coraggio[b], Carmela Calabrese[b], Agostino Guarino[b], Ricardo Cardona[b], Piero De Lellis[+, b], Davide Liuzza[+,c], Francesco Lo Iudice[+,b], Giovanni Russo[+,d], Mario di Bernardo[b,#]

[a]Department of Electronic, Information and Biomedical Engineering, Politecnico di Milano, Italy

[b]Department of Electrical Engineering and Information Technology, University of Naples Federico II, Italy

[c]ENEA, Fusion and Nuclear Safety Department, Frascati (Rome), Italy

[d]Department of Information and Electrical Engineering and Applied Mathematics, University of Salerno, Italy

*These authors contributed equally*

*+ These authors contributed equally*

# Correspondence: mario.di.bernardo@unina.it

**Index**

# S1. DATA ANALYSIS

*Data on the evolution of the epidemic in Italy*

Data on the evolution of the epidemic in Italy and in each of the 20 regions have been obtained from the public database[S7] of the Italian Civil Protection Agency (Protezione Civile). The database provides daily updates for each region on the overall number of detected cases, hospitalized, quarantined, and recovered. The data were pre-processed by filtering it with a moving average filter of 3 days to reduce noisiness. Information on the number of beds available in ICU in each region have been obtained from data reported in public repositories.[S8,S9]

*Interregional Fluxes estimation*

We considered two types of inter-regional fluxes associated to

(i) daily commuters traveling between neighboring regions;
(ii) long distance travels covered by high-speed trains, planes, and large ferries.

To estimate commuters' fluxes, as in previous work[S6], we use the latest official country-wide assessment of Italian mobility conducted by the Italian Institute of Statistic (ISTAT) in 2011. Specifically, we use the origin-destination matrix available at https://www.istat.it/it/archivio/139381 describing, at the municipality level, the number of people who declared themselves daily commuters for work or studying purposes. By aggregating data on a regional basis, we obtained interregional commuters' flows among regions.

For longer distance routes covered by high-speed trains, planes and large ferries we proceeded as follows.

a. For railway connections, the fraction $\phi_{ij}^r$ of the population of region i traveling to region *j* is obtained as

$$\phi_{ij}^r = \frac{1}{p_i}\frac{\phi^r}{n^r}n_{ij}^r,$$

where

- $\phi^r$ is total number of daily high-speed customers of the main Italian carrier, Trenitalia;
- $n^r$ is the total number of high-speed trains per day;
- $n^r_{ij}$ is the number of high-speed trains connecting region $i$ to region $j$;
- $p_i$ is the total resident population in region $i$.

b. For flight connections, for all the Italian operating airports (currently there are 39 according to www.enac.gov.it) and for each pair of connected airports, say $(l,k)$ the number of *weekly* direct flights $f_{lk}$ in normal operating conditions (i.e., without any restriction due to the emergency). Then, for each pair $(l,k)$ we computed the average capacity $\langle c_{lk} \rangle$ of the regional fleet of the main carrier serving the route. Grouping the airports on a regional basis we compute $\langle c_{lk} \rangle$ average daily flow due to air connections $\phi^a_{ij}$, from region $i$ to $j$ as:

$$\phi^a_{ij} = \frac{\sum_{l=1}^{N_i} \sum_{k=1}^{N_j} f_{lk} \langle c_{lk} \rangle}{7 p_i},$$

where $N_i$ and $p_i$ are the number of airports and the population of region $i$.

c. For ferry connections, we considered the five regions that act as hubs for long range national maritime travel, that is, the two main insular regions, Sardinia and Sicily, together with three mainland regions, Campania, Lazio, and Liguria. The maritime flows are then obtained as

$$\phi^m_{ij} = \frac{\phi_i}{p_i \, n^m_{ij}},$$

where $\phi_i$ and $p_i$ are the average number of maritime passengers and the population of region $i$ respectively, while $n^m_{ij}$ is the total number of maritime connections between regions $i$ and $j$. Note that as it is reasonable to assume that $n_i \phi_{ij} \approx n_j \phi_{ji}$, and as all relevant maritime routes are either from or to the main islands, in practice, it suffices to compute $\phi^m_{ij}$ only for region $i$ being Sicily and Sardinia.

## S2. IDENTITICATION PROCEDURE

As explained in the main text, we assume that, for each region, the parameters of the model remain constant over $n$ time windows, but neither their number $n$ nor their durations $\delta_1, \dots, \delta_n$ are assumed to be known *a priori*. Therefore, the identification procedure detects at the same time the breakpoints $t_1, \ldots, t_{n-1}$ when notable parameters' changes are detected and, within each time-window, estimates their values that best capture the trend of the available data.

The model used to carry out the identification of regional or national parameters is to start with the discretized version of the model of the epidemic spread in each area of interest given by model (14)-(20) in the main text.

As also noted in other previous work[S1-S4], identification of SIR and SIR-modified models is highly non-convex and hence the optimization landscape is scattered with local minima that must be avoided as not being admissible. To mitigate this problem, we identified from the literature admissible intervals for the parameter values (see Table S5) and fixed from the available evidence in the literature[S3] the parameters $\beta = 0.4$ and $\gamma = 1/14$.

Altogether, the unknown parameters left to be estimated are $[I(0), \rho, \tau, \alpha, \eta^Q, \eta^H, \kappa^H, \kappa^Q, \zeta]$ both at the national and regional level.

As mentioned in the main text, the identification procedure is carried out in two stages by considering equations (14)-(16) of the main text to estimate $\{\tau, I(0), \rho\}$ and the window breakpoints and equations (17)-(20) of the main text to estimate the remaining parameters. The identification described next is repeated for each of the 20 regions and, for the sake of completeness, to parameterize a national aggregate model.

*Step 1: Online Identification of the estimation breakpoints and the parameters $\tau, I(0), \rho$ in each time window*

We start by identifying the parameters' vector $\theta \coloneqq [I(0), \rho, \tau]$ exploiting equations (14)-(16) of the main text and the time series of the number of cases $\tilde{C}$ collected for $T_{\text{tot}}$ consecutive days starting from the day when 10 deceased and 10 recovered were first reported in the area of interest. In particular, an *ad hoc* optimization algorithm (described below and implemented in MATLAB) is used to find breakpoints $t_j$ and the values of the parameters' vector $\hat{\theta}$ that minimize the cumulative squared prediction error in each window, defined as

$$SSE(\hat{\theta}, t_j, t_{j+1}) = \sum_{t=t_j}^{t_{j+1}} \| \tilde{C}(t, \theta) - \hat{C}(t, \hat{\theta}) \|^2$$

with $j = 0, 1, 2, \ldots, n-1$.

We use the following recursive procedure:

1. Set the initial time $t_0$ as the first day in which the first 10 deaths and 10 recovered were reported in the area (region or nation).
2. Assume the initial guess for the width of the window to be $T = \lceil (2p+1)/d \rceil$, where $p$ is the number of parameters to be identified and $d$ the number of measured variables.
3. Estimate the parameters over the entire window to obtain $\hat{\theta} = \underset{\hat{\theta}}{\operatorname{argmin}} SSE(\hat{\theta}, t_0, t_0 + T)$
4. Divide the window into two intervals and estimate the parameters over each subsample obtaining the two estimates $\hat{\theta}_a = \underset{\hat{\theta}_a}{\operatorname{argmin}} SSE(\hat{\theta}_a, t_0, t_0 + \lceil T/2 \rceil)$, and $\hat{\theta}_b = \underset{\hat{\theta}_b}{\operatorname{argmin}} SSE(\hat{\theta}_b, t_0 + \lceil T/2 \rceil, t_0 + T)$.
5. Perform Chow statistical test

$$F = \frac{(T - 2p)(\sigma - (\sigma_a + \sigma_b))}{p(\sigma_a - \sigma_b)} \sim \mathcal{F}_{\{p, T-2p\}}$$

    where

$$\sigma = SSE(\hat{\theta}, t_0, t_0 + T)$$
$$\sigma_a = SSE(\hat{\theta}_a, t_0, t_0 + \lceil T/2 \rceil)$$
$$\sigma_b = SSE(\hat{\theta}_b, t_0 + \lceil T/2 \rceil, t_0 + T)$$

    and perform the test with null hypothesis $H_0: \{\hat{\theta}_1 = \hat{\theta}_2\}$ and critical $p$-value $p^* = 10^{-4}$

6.
    a. If $\mathcal{F}_{\{p, T-2p\}}(F) > p^*$, the null hypothesis cannot be rejected, and the parameters are considered constant in the time-window $T$. Then, the length of the current window is increased by setting $T = T + 1$, and steps 3, 4 and 5 are repeated.
    b. If $\mathcal{F}_{\{p, T-2p\}}(F) \leq p^*$, the null hypothesis is rejected, and then the next breakpoint $t_1$ is selected as

$$t_1 = \operatorname{argmax} \mathcal{F}_{\{p, T-2p\}}(F)$$

and the parameter set $\hat{\theta}$ that minimizes $SSE(\hat{\theta}, t_0, t_1)$ is selected as the set that best fits the data over the window $(t_0, t_1)$ whose duration is therefore $\delta_1 := t_1 - t_0$.

7. If $t_1 = T_{\text{tot}}$ the algorithm is stopped, otherwise starting from $t_1$ steps 2-7 are repeated to find the next breakpoint and the new set of parameters best estimating the data in the next window until the end of the available datapoints.

*Step 2: Offline refinement of the identification process*

At the end of the process we will have the set of breakpoints $t_j$ and the parameters set in each of the windows $(t_j, t_{j+1})$ best fitting the data. As the number of windows can be large given The variability in the available data, we refine the estimation results as follows to estimate the minimal number of windows able to capture qualitatively the trend of the real data.

In particular, once the window breakpoints are obtained at the end of step 1, any two consecutive windows of duration say $\delta_j, \delta_{j+1}$ are merged into one larger window of size $\delta_j + \delta_{j+1}$ if one of the two following conditions is verified:

a) The window size of the first window is less than 5 days.
b) The relative variation of the sum $\tau + \rho$ as estimated in each of the two windows is less than 5%, i.e.

$$\frac{(\tau + \rho)^{j+1} - (\tau + \rho)^j}{(\tau + \rho)^j} \leq 0.05$$

where *j* denotes the window to which the parameter estimates refers to.

If two windows are merged, then the parameters are estimated again on the entire merged window and the procedure is iterated once more in case condition b) is still verified.

As a final refinement step, we heuristically explore the effect on the fitting of perturbing the breakpoints within ±5 days from their estimated value. As a representative example, the results of the fitting procedure for the national aggregate model are shown in Table S3 and depicted in
Figure **S9**. The same procedure is repeated to parametrize each of the 20 regional models.

To provide a representative validation of our estimation approach, we report in Figure S10 the time evolution of the total number of detected cases at the national level predicted by the model in each time window. It is possible to see that using just 30% of the datapoints (shown in green) from all the available data (shown in red), the model predictions (solid blue lines) fit well the rest of the data in each time window both before and after the windows are merged as a result of step 2 with a maximum prediction error of 10000 units.

*Step 3: Identifying the parameters $\eta^H, \eta^Q, \psi, \alpha, \kappa^H, \kappa^Q, \zeta$*

For each of the time windows identified in Step 2, using the time series $\hat{I}(t)$ estimated from the equations parametrized in Step 1, and considering that equations (17)-(20) of the predictor reported in the main text are linear with respect to the parameters, we use a ordinary constrained least squares method, with constraints given in
**Table S2** to compute the remaining parameters.
The comparison between the model predictions and the available data is depicted in
Figure **S11** and
Figure **S12**. Values of all estimated parameters at the end of the process are given in
Table S4 [Model parameters]. Values of estimated parameters for each region at the end of the identification process. Dates are given corresponding to breakpoints between estimation windows that are automatically detected by the estimation procedure we proposed. for each region. In the last column the regional net reproduction numbers are computed as $R_{0,i} = \rho_i \beta / (\alpha_i + \psi_i + \gamma)$ in each time window. Figure S13 shows the distribution of the regional social distancing parameters over time showing the effects of the national lockdown at the regional level.

*$\zeta_i$ as a function of the occupancy of ICU beds in each region*

Observing the data and the identified parameters in
Table **S4**, we noticed a significant correlation between the mortality rate in each time window and the congestion of the ICU system in that region.

Specifically, we found that

$$\zeta_i = f(\bar{H}_i) = \zeta_0^{IC} + \zeta_1^{IC} \bar{H}_i$$

where $\zeta_0^{IC}$ and $\zeta_1^{IC}$ are coefficients to be estimated, while $\bar{H}_i$ is the estimated average congestion of the hospitals in each time-window defined as the ratio between the number of hospitalized people and the number of available beds in ICU in that region, say $T_i^H$ (available from the web-page of the Italian Ministry of Health[S8] and updated daily from the news published by the main Italian newspapers in each region).

Each point in Figure S14 is the value $\zeta_i$ estimated for each region in each time window against the number of hospitalized in the corresponding region averaged in the time window. As illustrated in Figure S14, a least square linear fitting yields $\zeta_0^{IC} = 0.016, \zeta_1^{IC} = 0.0013$.

## S3. REGIONAL FEEDBACK INTERVENTION STRATEGIES

We considered three different types of intervention strategies at the regional level that can be used individually or in combination.

1) *Feedback Social distancing rule.* Each region modulates its lockdown measures so as to switch them on or off according to the relative saturation level of its health system. Namely, the social distancing parameter is modulated as follows:

$$\rho_i = \begin{cases} \underline{\rho_i}, & if \ \frac{\widetilde{H}_i}{T_i^H} \geq 0.5 \\ \bar{\rho}_i, & if \ \frac{\widetilde{H}_i}{T_i^H} \leq 0.2 \end{cases}$$

where $\underline{\rho_i}$ is set equal to the minimum estimated value in that region during the national lockdown (corresponding to the value given in the last window for each region reported in Tab. S4) and, as a worst case scenario, $\bar{\rho}_i$ is set to three times $\underline{\rho_i}$ (or unity if $3\underline{\rho_i} > 1$) to simulate a relaxation of the containment measures in that region.

2) *Feedback flux control.* Here the fluxes in or out of a region are modulated according to the following rule

$$\phi_{ij} = \begin{cases} \underline{\phi}_{ij}, & if \ \frac{\widetilde{H}_i}{T_i^H} \geq 0.5 \\ \bar{\phi}_{ij}, & if \ \frac{\widetilde{H}_i}{T_i^H} \leq 0.2 \end{cases}, \forall j \neq i$$

and

$$\phi_{ji} = \begin{cases} \underline{\phi}_{ji}, & if \ \frac{\widetilde{H}_i}{T_i^H} \geq 0.5 \\ \bar{\phi}_{ji}, & if \ \frac{\widetilde{H}_i}{T_i^H} \leq 0.2 \end{cases}, \forall j \neq i$$

where we denote by $\underline{\phi}_{ij}$ and $\bar{\phi}_{ij}$ respectively the quarantine (low) and post quarantine (high) values of the flux from region $i$ to region $j$. In particular during a lockdown we set $\underline{\phi}_{ij} = 0.7\bar{\phi}_{ij}$ while $\bar{\phi}_{ij}$ correspond to the fluxes from region $i$ to region $j$ estimated as described in Sec. S.2.

Note that all people resident in $i$ not commuting to any other region are assumed to stay and move in region $i$ itself by setting $\phi_{ii} = 1 - \sum_{j \neq i} \phi_{ij}$.

**S4. FURTHER DETAILS ON THE NUMERICAL CODE**

All the numerical analyses presented in the paper were performed with Matlab. The code is available at https://github.com/diBernardoGroup/Network-model-of-the-COVID-19 and was designed to

1) Load the estimated parameter values and the inter-regional fluxes and to iterate the discretized model dynamics of each region considering the presence of interregional fluxes (MATLAB script "siqhrd_network_main.m"). The script also computes the regional and national net reproduction numbers.

2) Implement the differentiated regional intervention strategies described in Section S4.

3) Carry out parameter sensitivity analysis by using the Latin Hypercube sampling to explore the parameter region surrounding the estimated nominal parameter values. A variation of up to 20% of all parameters was considered in the simulations reported in the paper (Matlab scripts "siqrhd_network_main_montecarlo.m" and "hypercube_gen.m")

Further details can be found in the accompanying README.TXT file included with the code in the software repository above.

**SUPPLEMENTARY REFERENCES**

[S1] Mummert, A., & Otunuga, O. M. Parameter Identification for a Stochastic SEIRS Epidemic Model: Case Study Influenza. *Journal of Mathematical Biology* **79**, 705–729 (2019).

[S2] Liu, Y., Gayle, A. A., Wilder-Smith, A., & Rocklöv, J. The Reproductive Number of COVID-19 Is Higher Compared to SARS Coronavirus. *Journal of Travel Medicine* **27**, taaa021 (2020).

[S3] Giordano, G., Blanchini, F., Bruno, R., Colaneri, P., Di Filippo, A., Di Matteo, A., & Colaneri, M. Modelling the COVID-19 Epidemic and Implementation of Population-wide Interventions in Italy. *Nature Medicine*, 1–6 (2020).

[S4] Calafiore, G. C., Novara, C., & Possieri, C. A Modified SIR Model for the COVID-19 Contagion in Italy. Preprint at (2020).https://arxiv.org/abs/2003.14391 (2020).

[S5] Diekmann, O., Heesterbeek, J. A., & Metz, J. A. On the Definition and the Computation of the Basic Reproduction Ratio $R_0$ in Models for Infectious Diseases in Heterogeneous Populations. *Journal of Mathematical Biology* **28**, 365-382 (1990).

[S6] Gatto, M et al, Spread and dynamics of the COVID-19 Epidemic in Italy: Effects of Emergency Containment Measures. *Proceedings of the National Academy of Sciences* (2020)

[S7] https://github.com/pcm-dpc/COVID-19/tree/master/dati-regioni

[S8] http://www.dati.salute.gov.it/dati/dettaglioDataset.jsp?menu=dati&idPag=17

[S9] http://www.quotidianosanita.it/studi-e-analisi/articolo.php?articolo_id=82888.

[S10] Epicentro ISS, https://www.epicentro.iss.it/coronavirus/sars-cov-2-decessi-italia#2.

[S11] Epidemic Calculator, https://gabgoh.github.io/COVID/index.html.

[S12] N. M. Ferguson et al., "Impact of non-pharmaceutical interventions (NPIs) to reduce COVID- 19 mortality and healthcare demand" p. 20, 2020.

[S13] L. Peng, W. Yang, D. Zhang, C. Zhuge, and L. Hong, "Epidemic analysis of COVID-19 in China by dynamical modeling" Epidemiology, preprint, Feb. 2020.

[S14] F. Casella, "Can the COVID-19 epidemic be controlled on the basis of daily test reports?" arXiv:2003.06967, Apr. 2020.

[S15] M. Bin et al., "On Fast Multi-Shot Epidemic Interventions for Post Lock-Down Mitigation: Implications for Simple Covid-19 Models" arXiv:2003.09930, Apr. 2020.

[S16] S. Flaxman et al., "Estimating the number of infections and the impact of non-pharmaceutical interventions on COVID-19 in 11 European countries" 2020.

[S17] R. Li et al., "Substantial undocumented infection facilitates the rapid dissemination of novel coronavirus (SARS-CoV2)" Science, p. eabb3221, Mar. 2020.

[S18] R. Verity et al., "Estimates of the severity of coronavirus disease 2019: a model-based analysis" The Lancet Infectious Diseases, p. S1473309920302437, Mar. 2020.

[S19] http://github.com/pcm-dpc/COVID-19/tree/master/dati-andamento-nazionale.

**SUPPLEMENTARY FIGURES**

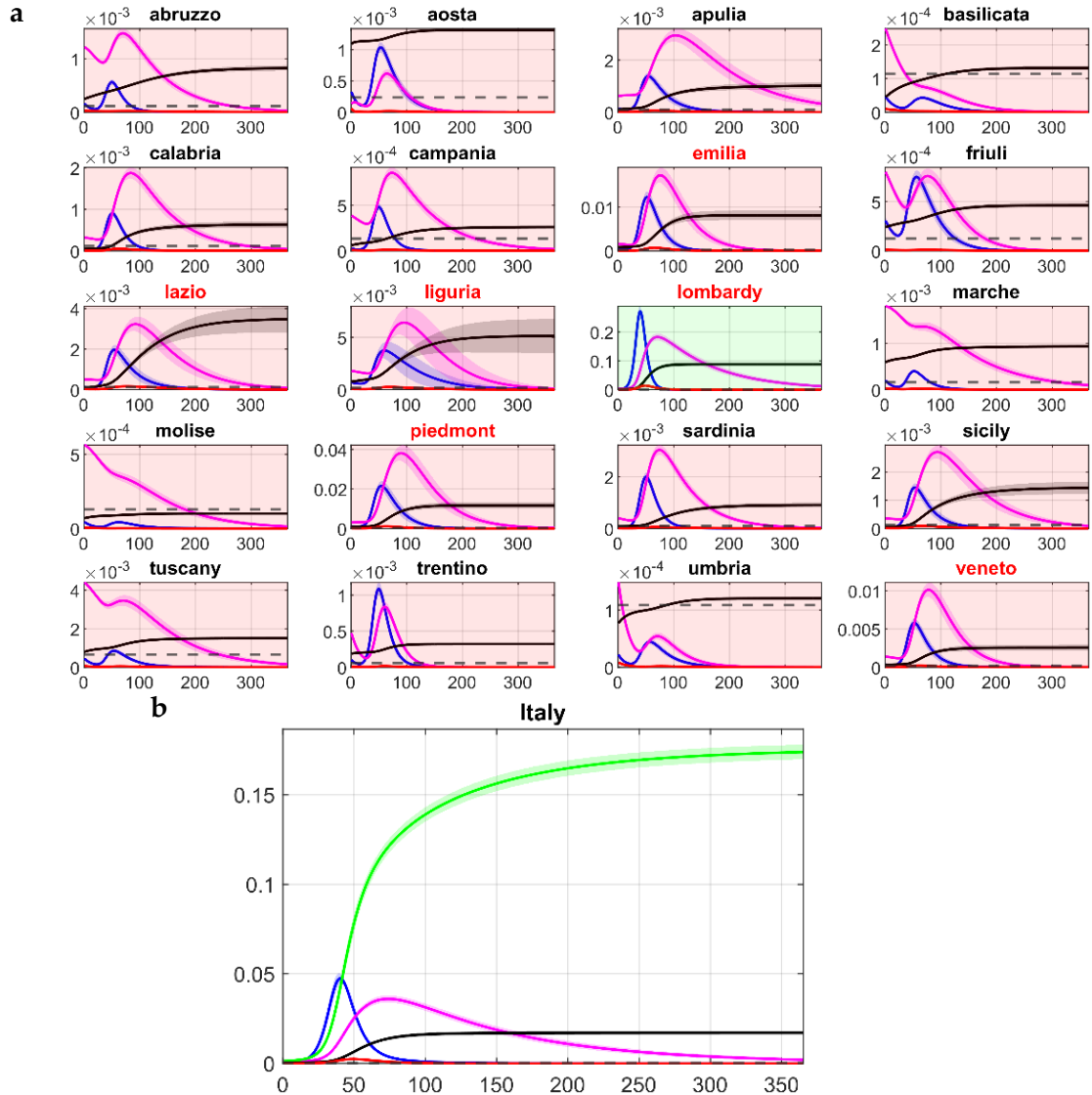

**Figure S1 [Phase 2, only one region relaxes its lockdown] a.** Regional and **b.** national dynamics in the case where only one region (Lombardy in Northern Italy) relaxes its containment measures at time 0 and the fluxes between regions are set to their pre-lockdown level. Panels of regions adopting a lockdown are shaded in red while those of regions relaxing social containment measures are shaded in green. blue, magenta, red, green, and black solid lines correspond to the fraction in the population of infected, quarantined, hospitalized, recovered, and deceased averaged over 10,000 simulations with parameters sampled using a Latin Hypercube technique around their nominal values set as those estimated in the last time window for each region as reported in
Table **S4**. Shaded bands correspond to twice the standard deviation. The black dashed line identifies the total fraction of the population that can be treated in ICU ($T_i^H/N_i$). The regions identified with a red label are those where the total hospital capacity is saturated.

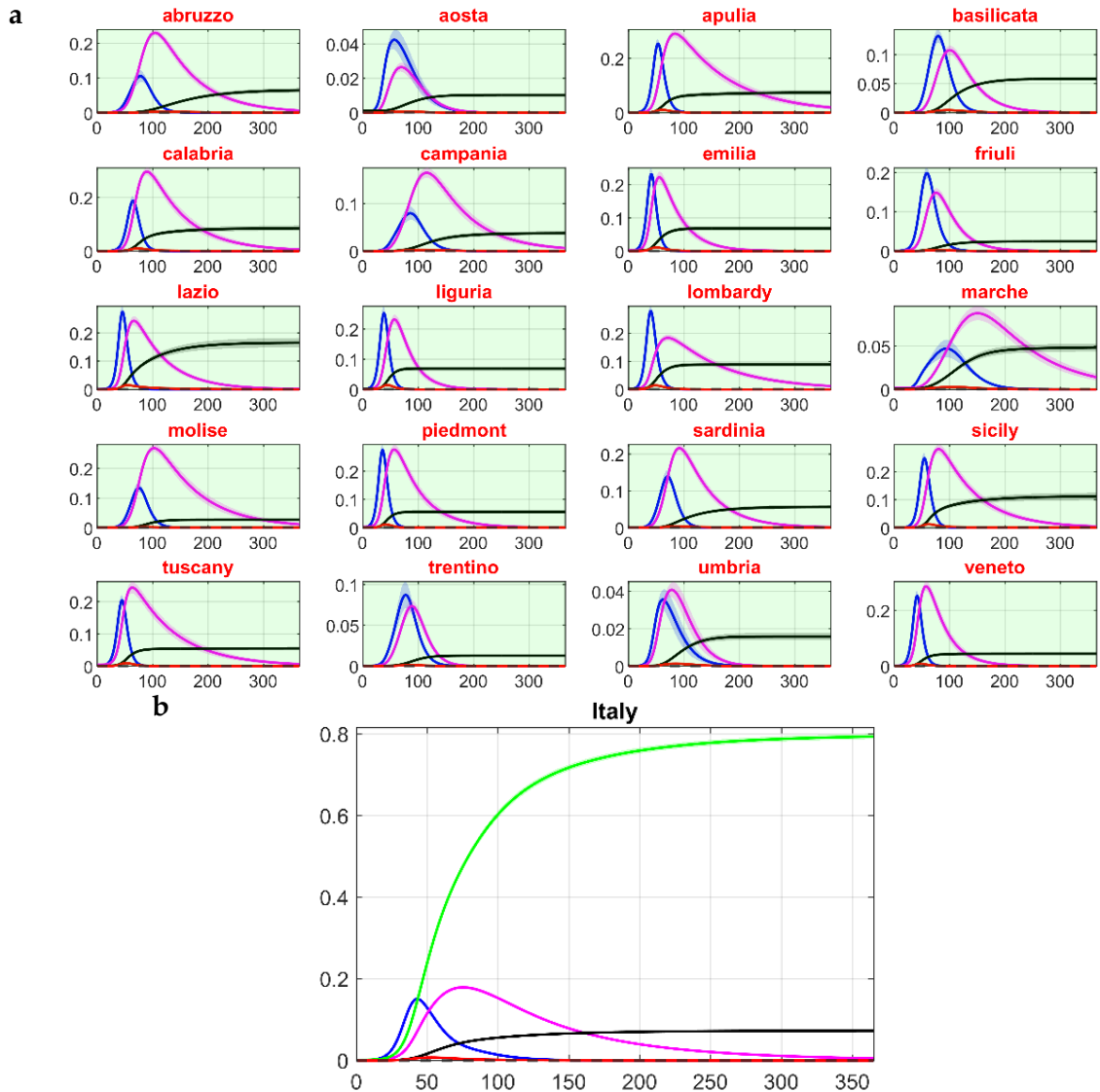

**Figure S2 [Phase 2, all regions relax their lockdown measures] ] a.** Regional and **b.** national dynamics in the case where all regions relax their current restrictions restoring fluxes to their pre-lockdown level. Blue, magenta, red, green, and black solid lines correspond to the fraction in the population of infected, quarantined, hospitalized, recovered, and deceased in the population averaged over 10,000 simulations with parameters sampled using a Latin Hypercube technique+ around their nominal values set as those estimated in the last time window for each region as reported in
Table **S4**. Shaded bands correspond to twice the standard deviation. The black dashed line identifies the fraction of the population that can be treated in ICU ($T_i^H/N_i$). The regions identified with a red label are those where the total hospital capacity is saturated.

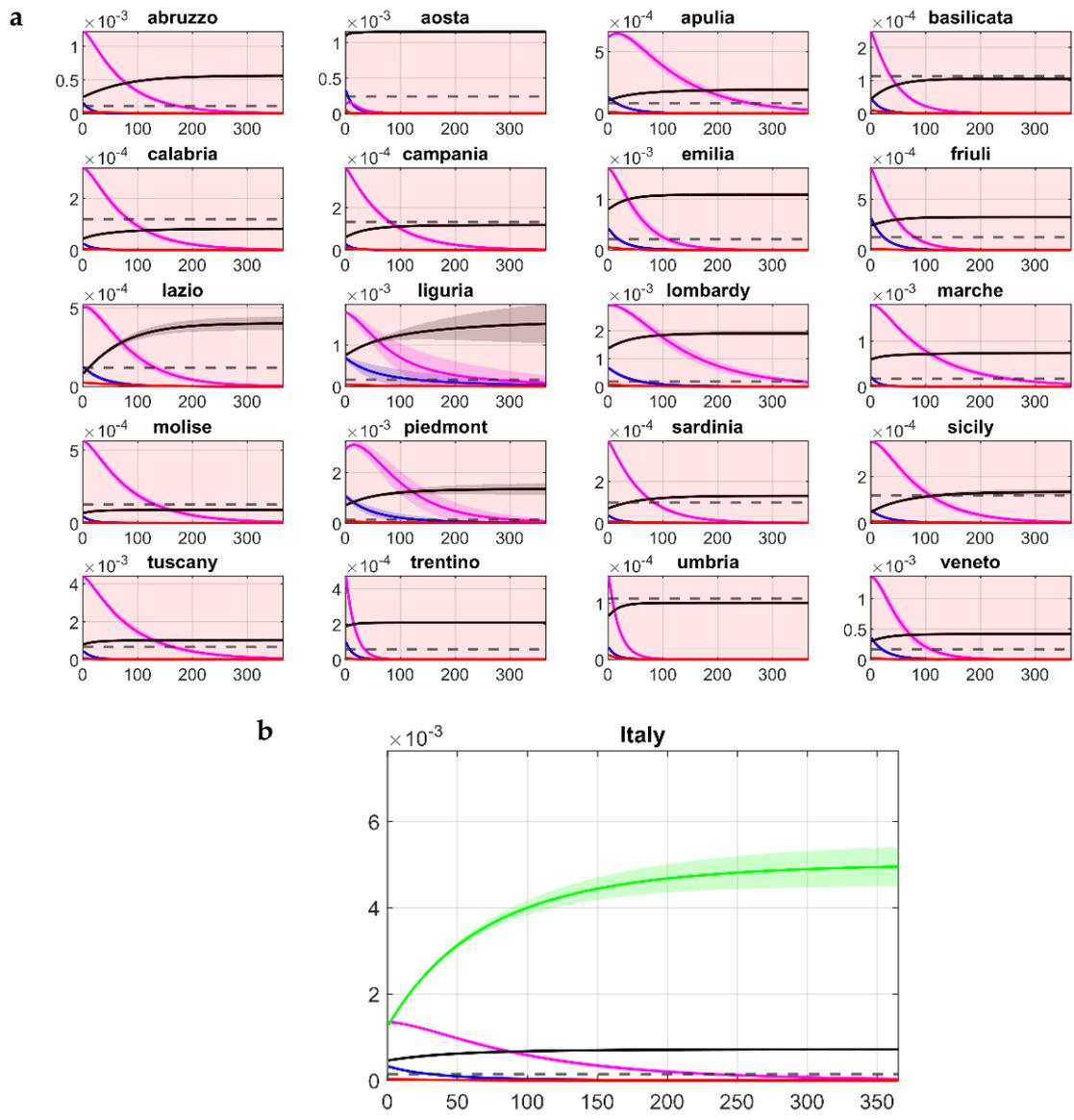

**Figure S3 [National lockdown]** **a.** Regional and **b.** national dynamics in the case where no region relaxes its containment measures, while all regions restore the interregional fluxes to their pre-lockdown level. Blue, magenta, red, green, and black solid lines correspond to the fraction in the population of infected, quarantined, hospitalized, recovered, and deceased averaged over 10,000 simulations with parameters sampled using a Latin Hypercube technique around their nominal values set as those estimated in the last time window for each region as reported in
Table **S4**. Shaded bands correspond to twice the standard deviation. The black dashed line identifies the fraction of the population that can be treated in ICU ($T_i^H/N_i$). The regions identified with a red label are those where the total hospital capacity is saturated.

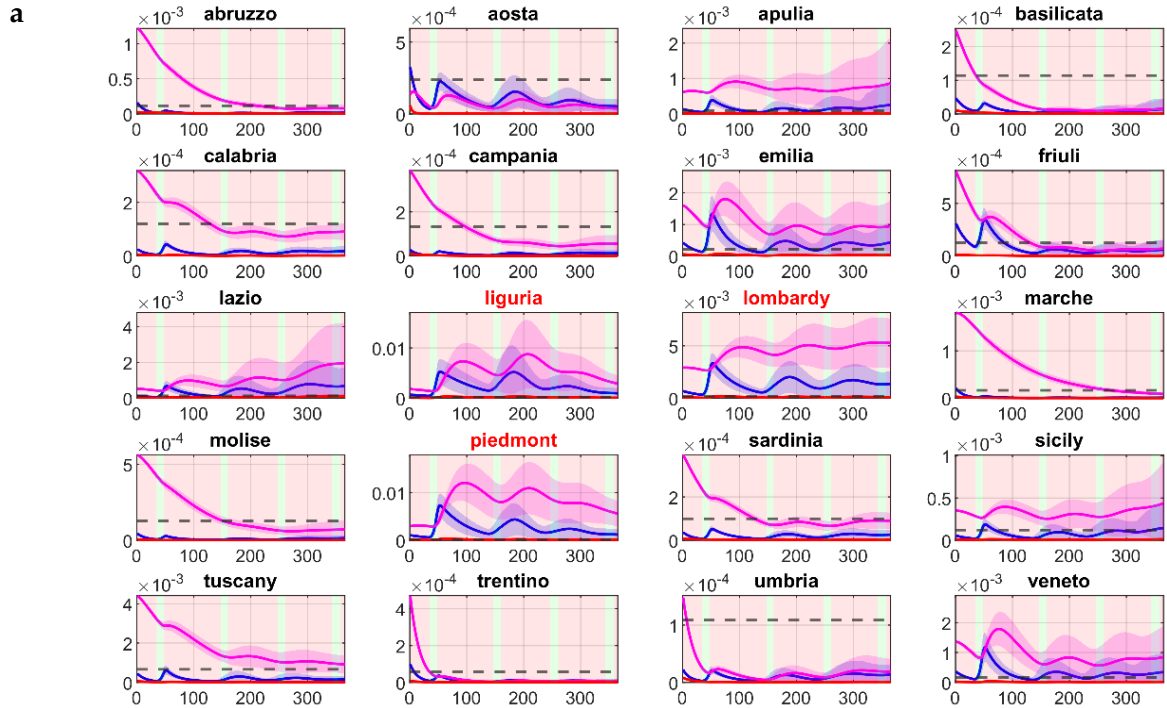

**Figure S4 [Phase 2, regional dynamics when intermittent national measures are enforced as shown in Fig. S3c of the main text].** Each of the 20 panels shows the evolution in the corresponding region of the fraction in the population of infected (blue), quarantined (magenta) and hospitalized requiring ICUs (red) averaged over 10,000 simulations with parameters sampled using a Latin Hypercube technique around their nominal values set as those estimated in the last time window for each region as reported in Table S9. Shaded bands correspond to twice the standard deviation. The black dashed line identifies the fraction of the population that can be treated in ICU ($T_i^H/N_i$). National lockdown measures are enforced with all regions shutting down when the total number of occupied ICU beds at the national level exceed 20% (windows shaded in red, green when relaxed). The regions identified with a red label are those where the total hospital capacity is saturated.

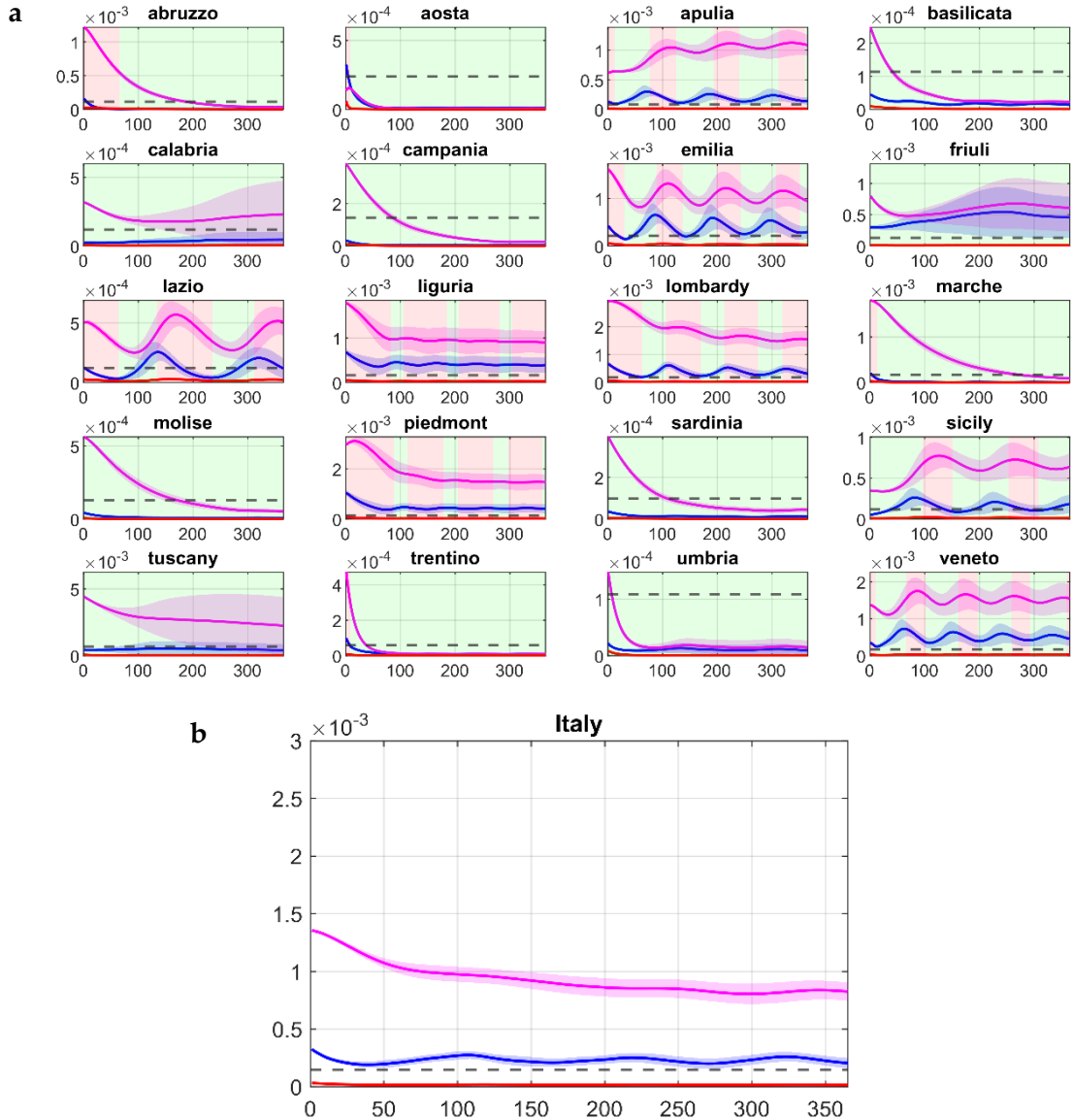

**Figure S5 [Phase 2, intermittent regional measures, $\bar{\rho}_i = 1.5\underline{\rho}_i$]. a.** Each of the 20 panels shows the evolution in the region named above the panel of the fraction in the population in the population of infected (blue), quarantined (magenta) and hospitalized requiring ICUs (red) averaged over 10,000 simulations with parameters sampled using a Latin Hypercube technique around their nominal values set as those estimated in the last time window for each region as reported in
Table **S4** of the Supplementary Information. Shaded bands correspond to twice the standard deviation. The black dashed line identifies the fraction of the population that can be treated in ICU ($T_i^H/N_i$). Regions adopt lockdown measures in the time windows shaded in red while relax them in those shaded in green. Differently from Figure 3, when lockdown measures are relaxed $\bar{\rho}_i$ is set to 1.5 times $\underline{\rho}_i$. During a regional lockdown, fluxes in/out of the region are set to their minimum level. **b.** National evolution of the fraction in the population of infected (blue), quarantined (magenta) and hospitalized requiring ICUs (red) obtained by summing those in each of the 20 regions adopting intermittent regional measures.

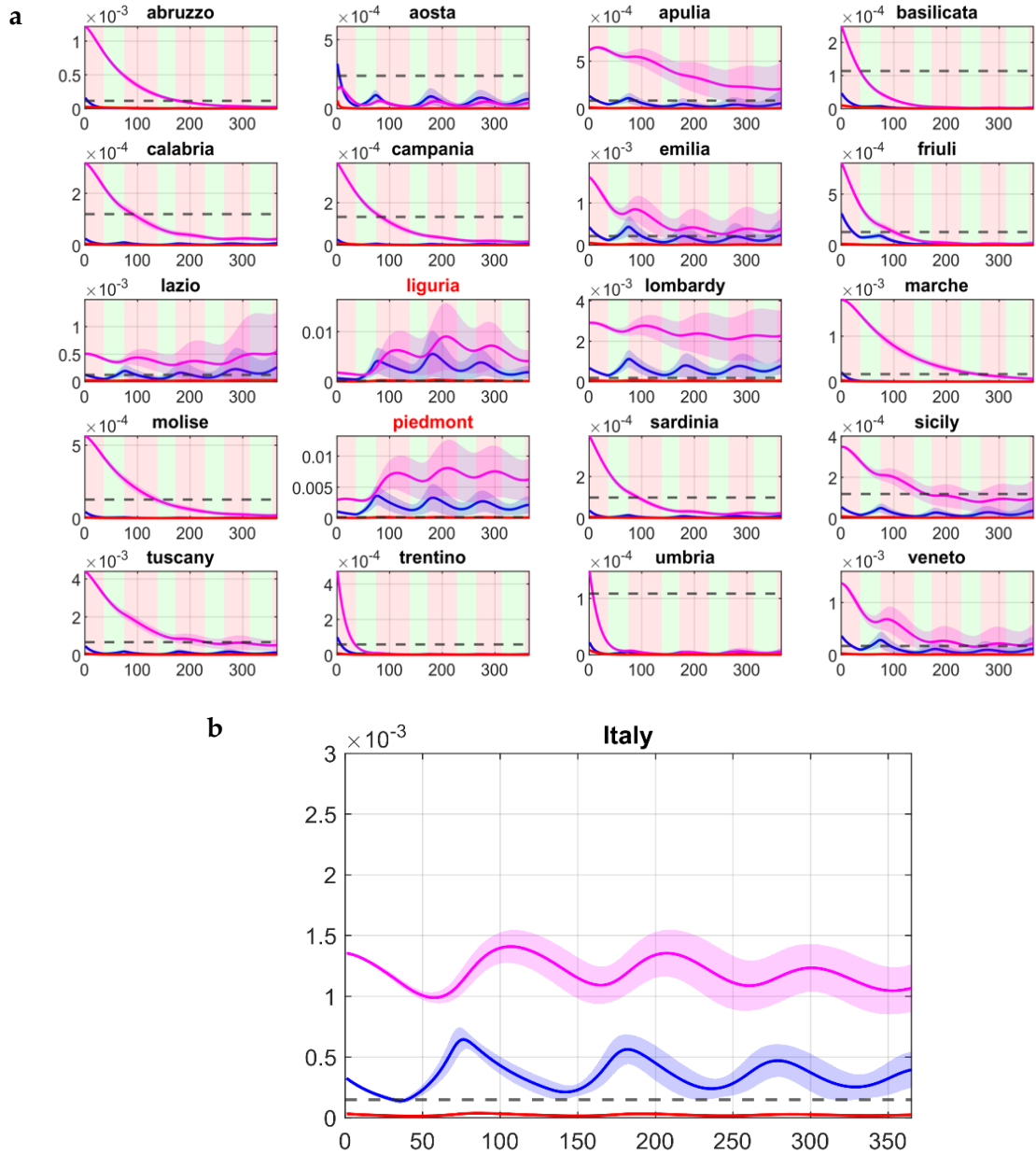

**Figure S6 [Phase 2, intermittent national measures, $\bar{\rho}_i = 1.5\underline{\rho}_i$] a.** Each of the 20 panels shows the evolution in the corresponding region of the fraction in the population of infected (blue), quarantined (magenta) and hospitalized requiring ICUs (red) averaged over 10,000 simulations with parameters sampled using a Latin Hypercube technique around their nominal values set as those estimated in the last time window for each region as reported in Table **S4**. Shaded bands correspond to twice the standard deviation. The black dashed line identifies the fraction of the population that can be treated in ICU ($T_i^H/N_i$). National lockdown measures are enforced with all regions shutting down when the total number of occupied ICU beds at the national level exceed 20%. The regions identified with a red label are those where the total hospital capacity is saturated. Regions adopt lockdown measures in the time windows shaded in red while relax them in those shaded in green. Differently from Figure S4, when lockdown measures are relaxed $\bar{\rho}_i$ is set to 1.5 times $\underline{\rho}_i$. During a regional lockdown, fluxes in/out of the region are set to their minimum level. **b.** National evolution of the fraction in the population of infected (blue), quarantined (magenta) and hospitalized requiring ICUs (red) obtained by summing those in each of the 20 regions.

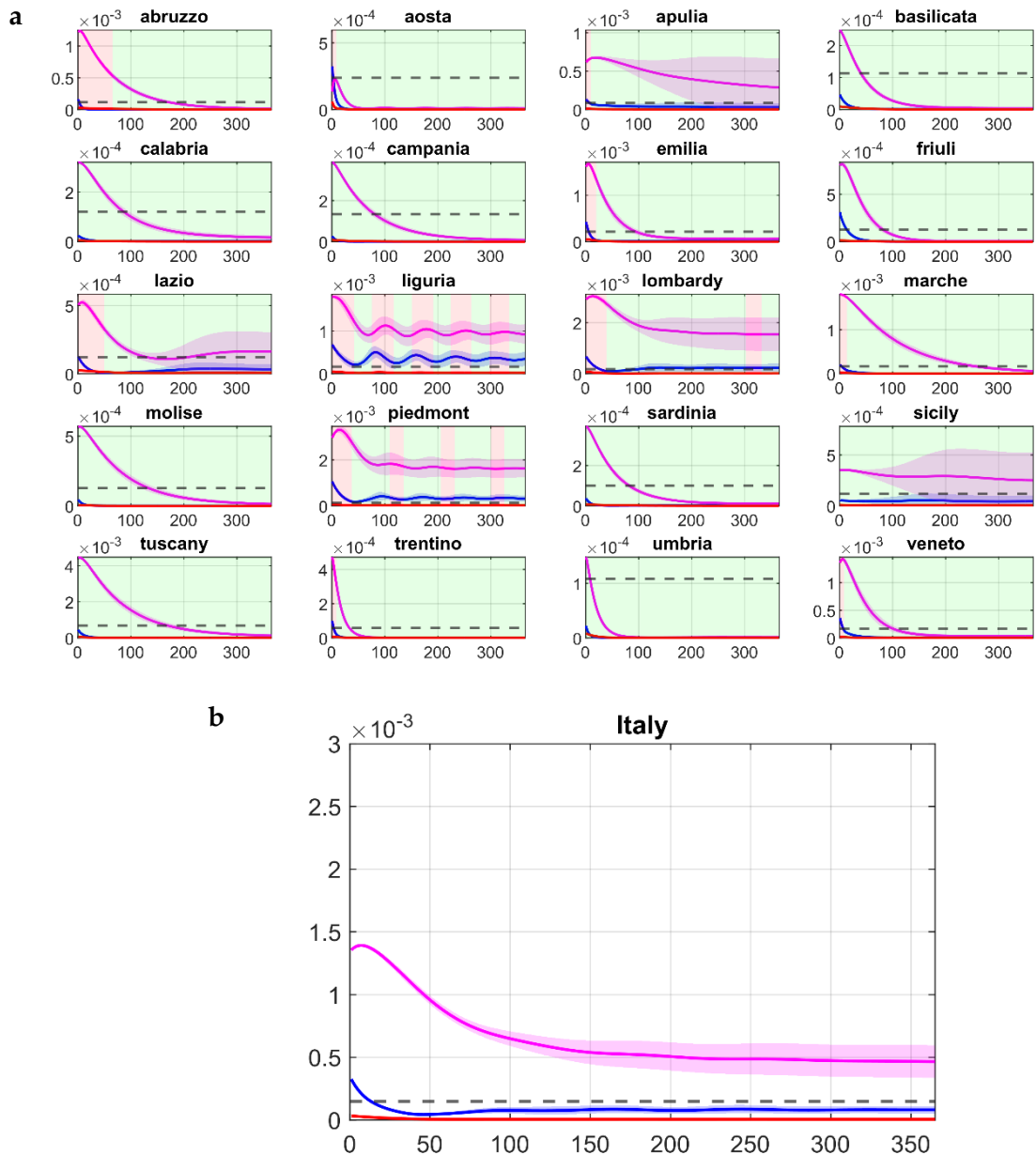

**Figure S7 [Phase 2, intermittent regional measures with increased COVID-19 testing capacity and $\bar{\rho}_i = 1.5\underline{\rho}_i$] a.** Each of the 20 panels shows the evolution in the region named above the panel of the fraction in the population of infected (blue), quarantined (magenta) and hospitalized requiring ICUs (red) averaged over 10,000 simulations with parameters sampled using a Latin Hypercube technique (see Methods) around their nominal values set as those estimated in the last time window for each region as reported in
Table **S4**. Shaded bands correspond to twice the standard deviation. The black dashed line identifies the fraction of the population that can be treated in ICU ($T_i^H/N_i$). Regions adopt lockdown measures in the time windows shaded in red while relax them in those shaded in green. During a regional lockdown, fluxes in/out of the region are set to their minimum level. Regions COVID-19 testing capacities are assumed to be increased by a factor 2.5 (see Methods) with respect to their current values. Differently from Figure 4, when lockdown measures are relaxed $\bar{\rho}_i$ is set to 1.5 times $\underline{\rho}_i$. During a regional lockdown, fluxes in/out of the region are set to their minimum level. **b.** National evolution of the fraction in the population of infected (blue), quarantined (magenta) and hospitalized requiring ICUs (red) obtained by summing those in each of the 20 regions.

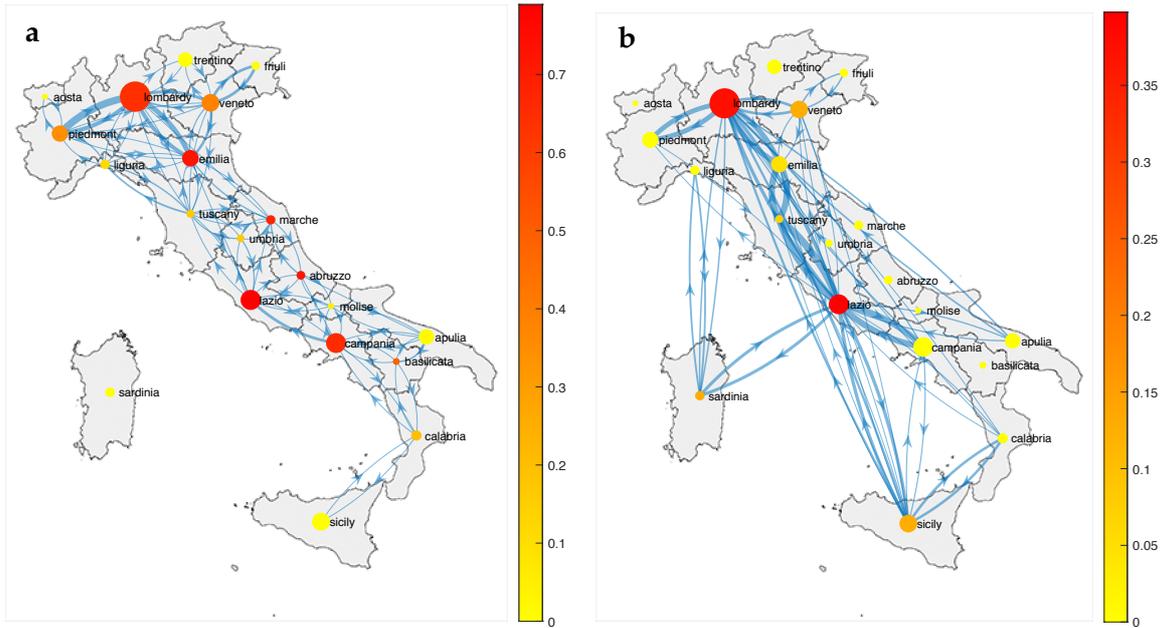

**Figure S8 [Networks resulting from the estimation of flows]** (a) Daily commuters' network; (b) Network topology resulting from the estimation of flows due to high-speed trains, planes, and ferry connections. For the sake of clarity, edges with neglible fluxes are not shown in the figure. Node colours are a measure of their betweenness centrality.

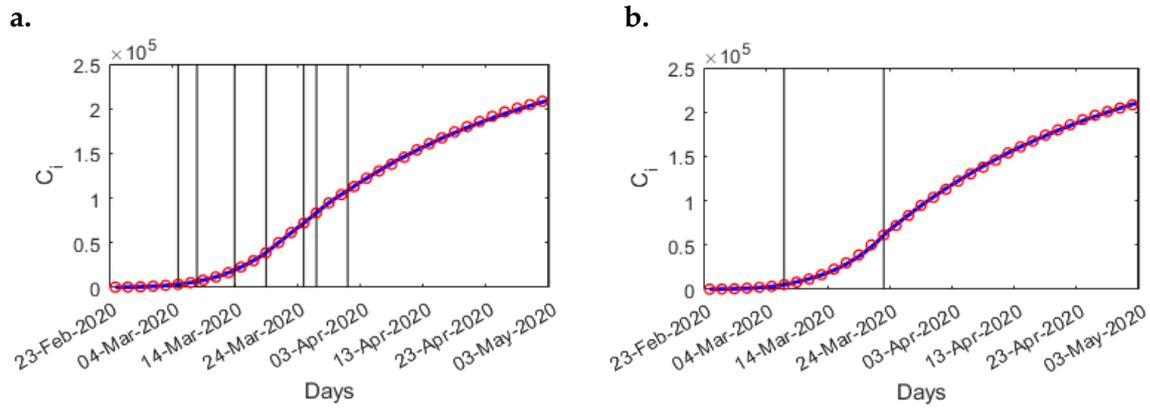

**Figure S9 [Identification of the aggregate national model]** Panel (a). Comparison between model predictions and data collected with time widows Identified at the end of Step 1 of the parameter identification process. Panel (b). Comparison between model predictions and data collected with the merged time widows obtained after step 2. In both panels the estimated number of cases predicted by the model $\hat{C}$ (blue solid line) is compared with the available datapoints $\tilde{C}$ (shown as red circles).

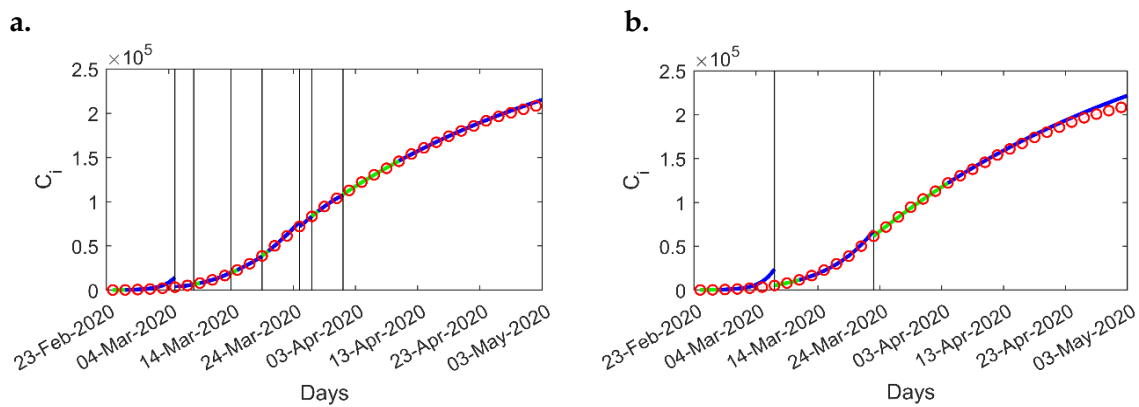

**Figure S10. [Example of data matching and prediction ability].** Estimation of the total number of detected cases $C_i$ (blue solid line) at the national level when the parameters are estimated using the 30% (green circles) of all the available data (red circles) in each time window. As shown in panel (a) at the end of step 1, the model predictions match well all the data point in each window. Panel (b) shows the predictive ability of the model after the windows are merged as a result of Step 2 of the parameter estimation process.

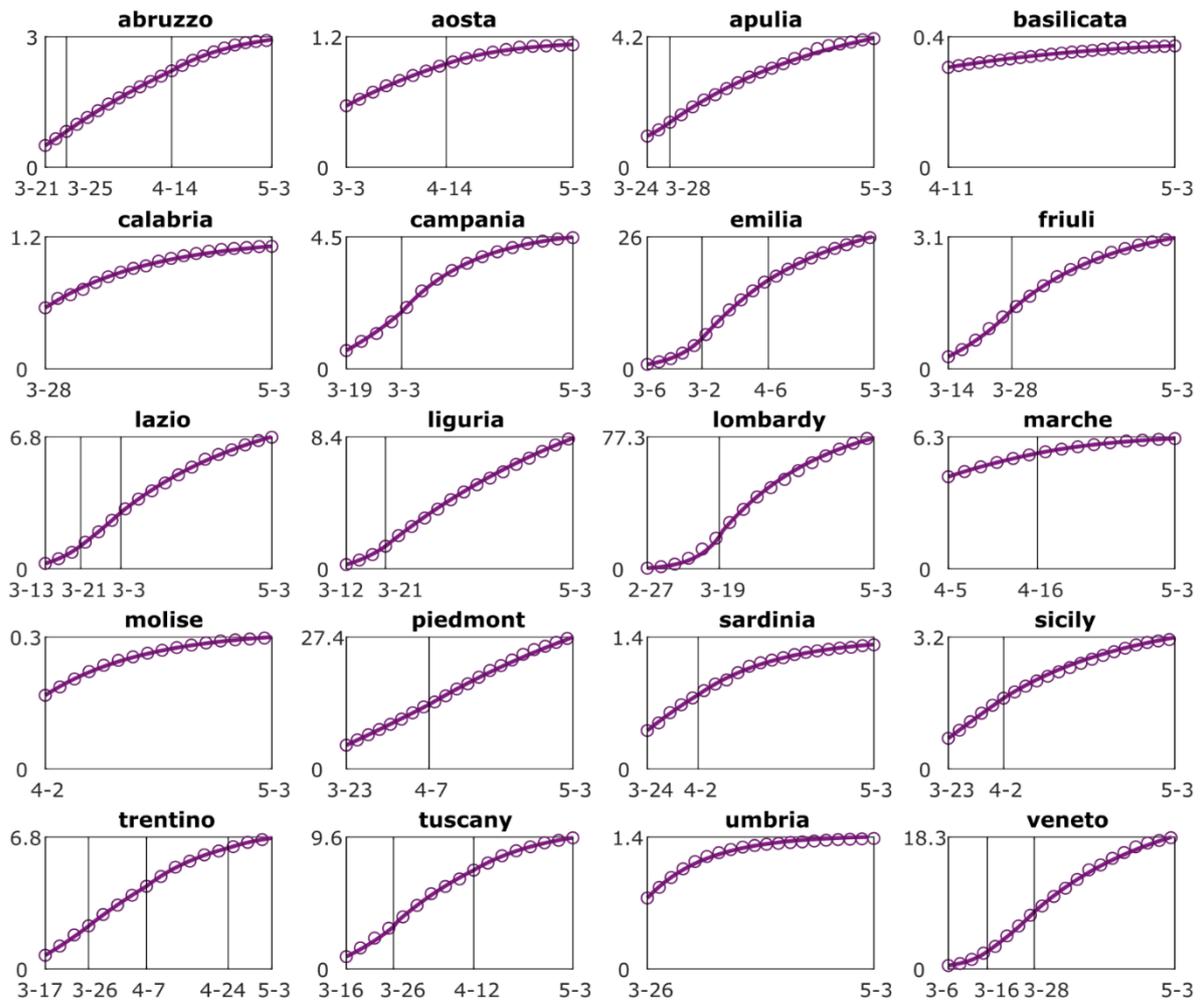

**Figure S11 [Identification of the regional models - Steps 1,2]** Comparison of each of the regional model predictions for the total number (expressed in thousands of people) of detected cases in each region (solid magenta line) against the available data points. Parameters are set to the values estimated at the end of Steps 2 carried out for each region. Vertical black lines denote the breakpoints from one estimation window to the next.

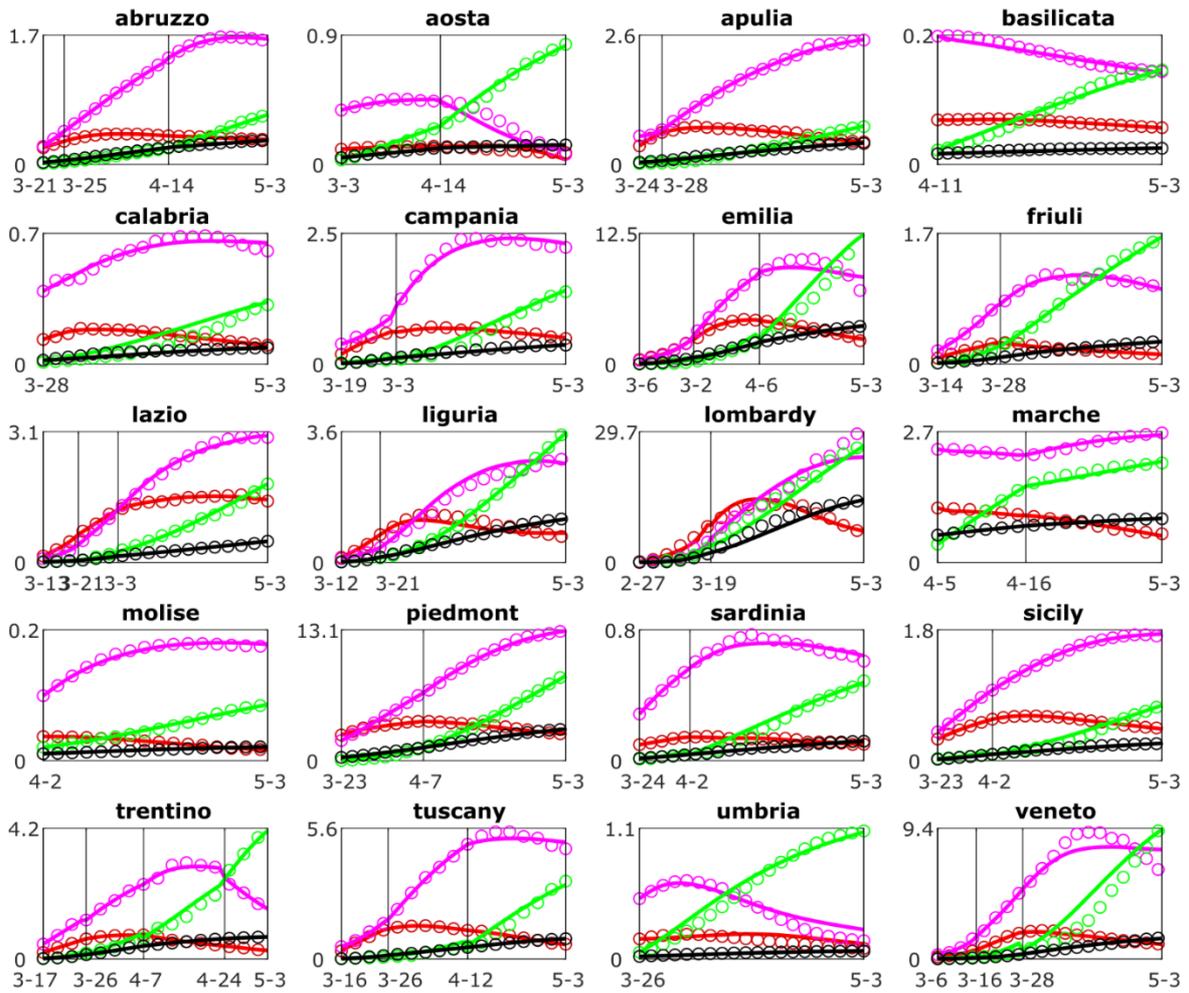

**Figure S12 [Identification of the regional models - Step 3]** Comparison of each of the regional model predictions of the total number (expressed in thousands of people) of recovered (green), quarantined (magenta), hospitalized (red), deceased (black) and recovered (green) in each region against the available data points (plotted as circles of the same colour). Parameters are set to the values estimated at the end of Step 3 carried out for each region. Vertical black lines denote the breakpoints from one estimation window to the next.

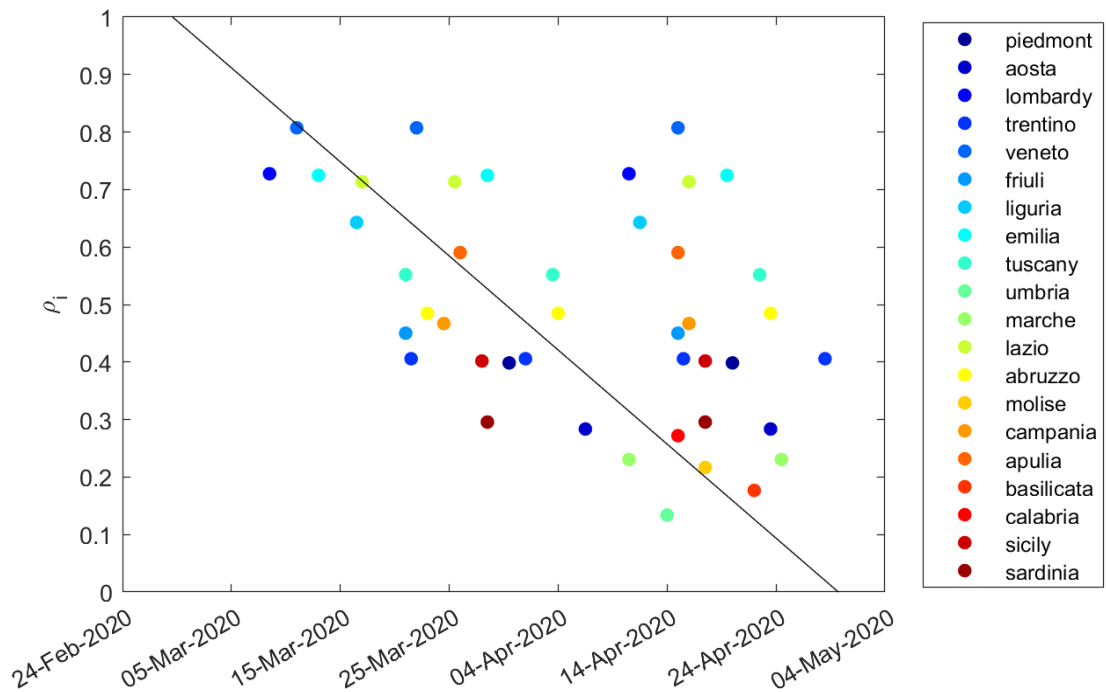

**Figure S13 [Distribution of the social distancing parameter over time]** Distribution of the social distancing parameter $\rho_i$ in the different detected estimation windows. The black line is the LS-interpolant of the National data reported in
Table S4

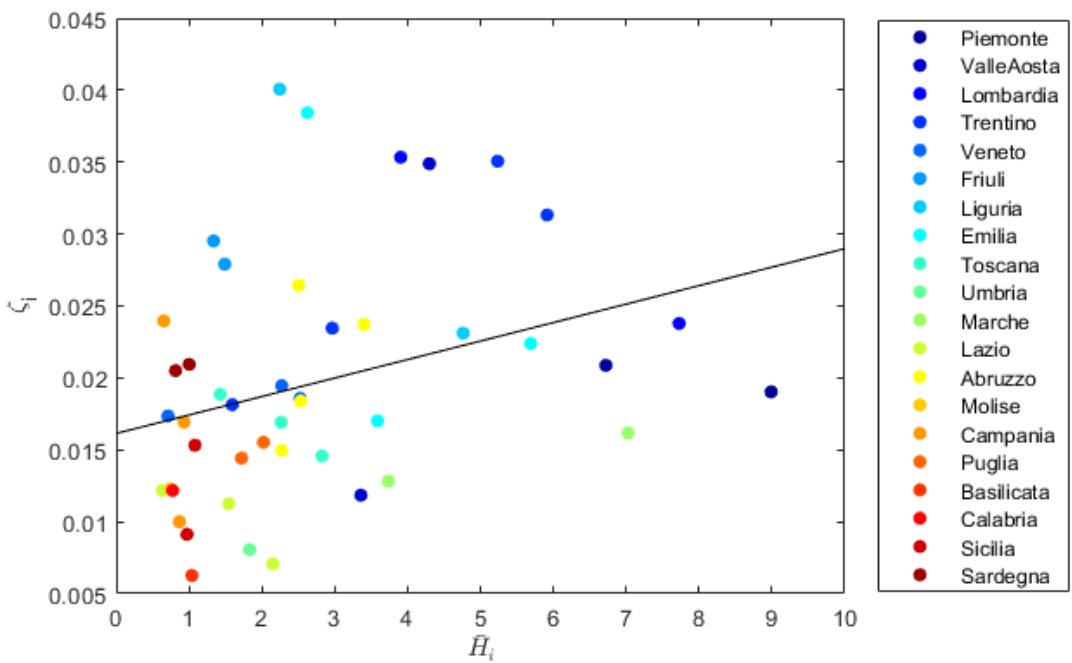

**Figure S14.** $\zeta_i$ fitting of $\bar{H}_i$ as a function of $R^2 = 0.280$ $p < 0.001$ and $p < 0.001$. Normality of the residuals has been tested with the Lilliefors test (p-value 0.12).

**SUPPLEMENTARY TABLES**

| Simulation scenario | Total cases | Total deaths | Maximum hospitalized | Days over hospital's capacity (nation) | Regions over hospital's capacity | Economic cost [M€] |
|---|---|---|---|---|---|---|
| All regions but Lombardy are locked down, Figure 2 | 10,545,380 ± 146,458 | 956,442 ± 68,215 | 144,180 ± 10,100 | 78.6 ± 2.3 | 3 | 475,330 ± 0 |
| Intermittent regional measures, Fig. 3a,b | 2,165,229 ± 83,806 | 154,878 ± 3,008 | 2,927 ± 183 | 0 ± 0 | 0 | 470,735 ± 6,353 |
| Intermittent national measure, Figure 3c, Figure S4 | 2,197,076 ± 189,948 | 176,210 ± 8,264 | 4,794 ± 309 | 0 ± 0 | 3 | 532,802 ± 12,474 |
| Intermittent regional measures with increased testing, Figure 4 | 939,091 ± 39,414 | 74,017 ± 1,325 | 1,915 ± 0 | 0 ± 0 | 0 | 349,963 ± 9,732 |
| All regions but Lombardy are locked down (pre-lockdown fluxes), Figure S1 | 11,650,960 ± 209,949 | 1,033,100 ± 66,786 | 147,316 ± 9,763 | 84.9 ± 2.3 | 6 | 475,330 ± 0 |
| No measure is taken, Figure S2 | 52,698,290 ± 355,425 | 4,417,306 ± 153,248 | 449,912 ± 15,857 | 220.2 ± 6.6 | 20 | 0 ± 0 |
| National lockdown, Figure S3 | 345,552 ± 29,841 | 43,705 ± 1,715 | 1,915 ± 0 | 0 ± 0 | 0 | 610,480 ± 0 |
| Intermittent regional measures $\bar{\rho}_i = 1.5\underline{\rho}_i$ Figure S5 | 1,023,858 ± 57,921 | 87,785 ± 2,394 | 1,915 ± 0 | 0 ± 0 | 0 | 245,121 ± 18,986 |
| Intermittent national measure Figure S6 | 1,409,581 ± 136,169 | 113,227 ± 3,720 | 2,270 ± 94 | 0 ± 0 | 2 | 375,082 ± 41,254 |
| Intermittent regional measures with increased testing and $\bar{\rho}_i = 1.5\rho_i$, Figure S7 | 542,560 ± 66,170 | 54,347 ± 4,500 | 1,915 ± 0 | 0 ± 0 | 0 | 70,995 ± 15,480 |

**Table S1 [Comparison of each of the simulated scenarios over 1 year].** Metrics to evaluate the impact over 1 year of each of the simulated scenarios are reported showing the effectiveness of the Intermittent regional measures in avoiding any saturation of the regional health systems while mitigating the impact of the epidemic and reducing the costs.

| Constraint | Description |
|---|---|
| $0.9\hat{I}^j \leq \hat{I}^{j+1} \leq 1.1\hat{I}^j$; | Continuity constraint on the number of infected at the national level from window $j$ to window $j+1$. |
| $LB \leq \hat{I}_i^{j+1} \leq UB$ <br> $LB = 0.9\hat{I}_i^j - 0.1\hat{I}^j$ <br> $UB = 1.1\hat{I}_i^j + 0.1\hat{I}^j$ | Continuity constraint on the number of infected at the regional level. The constraints are relaxed by 10% of the national estimated infected to account for the fact that in estimating the region parameter we are neglecting the influx of infected from other regions. |
| $\hat{\kappa}^Q \leq 0.1$, <br> $\hat{\kappa}^H \leq 0.1$; | We assume that the daily number of people hospitalized from quarantine and discharged but still positive (and vice versa) is no higher than 10% of the total. |
| $0.7\tau^j \leq \alpha^j + \psi^j \leq 1.3\tau^j$ | We assume $\hat{\tau}_i = \hat{\alpha}_i + \hat{\psi}_i$ does not differ from the national estimate $\hat{\tau}$ of more than 30%. |
| $\eta^{Q,j+1} = \eta^{Q,j}$ | We assume the recovery rate of those quarantined at home remains the same from a time window to the next as this parameter is likely to be time-invariant. In any case, removing this constraint, we observed no significant change of this parameter from a time window to the next. |

**Table S2 [Constraints for the regional models' parameterization]**. Set of parameters constraints enforced by the ordinary least square algorithm used for Step 2.

|  | $\rho$ | $I_0$ | $I_f$ | $\eta_Q$ | $\eta_H$ | $\zeta$ | $\alpha$ | $\psi$ | $\kappa_H$ | $\kappa_Q$ | $t_i$ | $R_{0,i}$ |
|---|---|---|---|---|---|---|---|---|---|---|---|---|
| | | | | | *Step 1* | | | | | | | |
| * | 0.965 | 1,200 | 11,580 | 0.028 | 0.045 | 0.023 | 0.020 | 0.050 | 0.000 | 0.000 | 24-Feb-20 | 2.76 |
| * | 0.958 | 11,580 | 18,805 | 0.028 | 0.013 | 0.024 | 0.018 | 0.059 | 0.027 | 0.049 | 05-Mar-20 | 2.61 |
| † | 0.774 | 18,805 | 43,814 | 0.028 | 0.017 | 0.029 | 0.034 | 0.042 | 0.000 | 0.000 | 08-Mar-20 | 2.11 |
| † | 0.607 | 43,814 | 65,971 | 0.028 | 0.019 | 0.029 | 0.048 | 0.025 | 0.100 | 0.000 | 14-Mar-20 | 1.70 |
| † | 0.331 | 65,971 | 56,726 | 0.028 | 0.008 | 0.032 | 0.053 | 0.037 | 0.099 | 0.100 | 19-Mar-20 | 0.83 |
| ‡ | 0.533 | 56,726 | 59,229 | 0.028 | 0.002 | 0.029 | 0.061 | 0.042 | 0.045 | 0.054 | 25-Mar-20 | 1.24 |
| ‡ | 0.234 | 59,229 | 44,109 | 0.028 | 0.000 | 0.027 | 0.049 | 0.043 | 0.056 | 0.100 | 27-Mar-20 | 0.58 |
| ‡ | 0.374 | 44,109 | 17,980 | 0.028 | 0.000 | 0.018 | 0.023 | 0.085 | 0.001 | 0.088 | 01-Apr-20 | 0.84 |
| | | | | | *Step 2* | | | | | | | |
| * | 0.937 | 1,200 | 16,639 | 0.028 | 0.031 | 0.023 | 0.010 | 0.058 | 0.000 | 0.000 | 24-Feb-20 | 2.70 |
| † | 0.646 | 16,639 | 78,621 | 0.028 | 0.014 | 0.032 | 0.018 | 0.072 | 0.000 | 0.100 | 07-Mar-20 | 1.62 |
| ‡ | 0.298 | 78,621 | 26,392 | 0.028 | 0.000 | 0.020 | 0.019 | 0.056 | 0.000 | 0.080 | 23-Mar-20 | 0.82 |

**Table S3 [Parameters of the aggregate national model]** Parameters of the aggregate national model before and after Step 2. Parameters values are given before and after merging the time windows. Symbols at the beginning of each row denote parameters from windows that are then merged in Step 2 of the identification procedure.

| Region | ρ | $I_0$ | $I_f$ | $\eta_Q$ | $\eta_H$ | ζ | α | ψ | $\kappa_H$ | $\kappa_Q$ | $t_i$ | $R_0$ |
|---|---|---|---|---|---|---|---|---|---|---|---|---|
| Abruzzo | 0,485 | 944 | 1083 | 0,010 | 0,003 | 0,026 | 0,029 | 0,051 | 0,005 | 0,099 | 21-mar-20 | 1,29 |
|  | 0,321 | 1083 | 747 | 0,010 | 0,000 | 0,022 | 0,025 | 0,049 | 0,000 | 0,087 | 25-mar-20 | 0,89 |
|  | 0,194 | 847 | 210 | 0,010 | 0,019 | 0,014 | 0,078 | 0,003 | 0,005 | 0,000 | 14-apr-20 | 0,51 |
| Aosta | 0,283 | 425 | 262 | 0,010 | 0,096 | 0,036 | 0,062 | 0,016 | 0,028 | 0,000 | 30-mar-20 | 0,77 |
|  | 0,122 | 262 | 41 | 0,010 | 0,260 | 0,011 | 0,062 | 0,000 | 0,079 | 0,000 | 14-apr-20 | 0,37 |
| Apulia | 0,590 | 1300 | 1732 | 0,010 | 0,000 | 0,016 | 0,028 | 0,047 | 0,100 | 0,100 | 24-mar-20 | 1,62 |
|  | 0,278 | 1732 | 530 | 0,010 | 0,004 | 0,015 | 0,022 | 0,051 | 0,002 | 0,078 | 28-mar-20 | 0,78 |
| Basilicata | 0,177 | 90 | 26 | 0,010 | 0,060 | 0,006 | 0,037 | 0,021 | 0,025 | 0,025 | 11-apr-20 | 0,55 |
| Calabria | 0,272 | 384 | 49 | 0,015 | 0,000 | 0,012 | 0,042 | 0,059 | 0,005 | 0,079 | 28-mar-20 | 0,63 |
| Campania | 0,467 | 1231 | 1816 | 0,018 | 0,000 | 0,022 | 0,014 | 0,064 | 0,000 | 0,100 | 19-mar-20 | 1,26 |
|  | 0,221 | 2231 | 234 | 0,018 | 0,000 | 0,011 | 0,067 | 0,019 | 0,006 | 0,040 | 30-mar-20 | 0,57 |
| Emilia | 0,725 | 1418 | 7246 | 0,029 | 0,000 | 0,038 | 0,020 | 0,089 | 0,000 | 0,100 | 06-mar-20 | 1,62 |
|  | 0,400 | 7246 | 4467 | 0,029 | 0,000 | 0,023 | 0,059 | 0,062 | 0,000 | 0,045 | 20-mar-20 | 0,84 |
|  | 0,362 | 4467 | 1881 | 0,029 | 0,031 | 0,017 | 0,063 | 0,050 | 0,000 | 0,017 | 06-apr-20 | 0,79 |
| Friuli | 0,450 | 900 | 1717 | 0,028 | 0,022 | 0,028 | 0,032 | 0,034 | 0,000 | 0,100 | 14-mar-20 | 1,32 |
|  | 0,202 | 1717 | 376 | 0,028 | 0,049 | 0,029 | 0,044 | 0,007 | 0,004 | 0,000 | 28-mar-20 | 0,67 |
| Lazio | 0,713 | 722 | 1689 | 0,015 | 0,003 | 0,013 | 0,018 | 0,081 | 0,000 | 0,062 | 13-mar-20 | 1,69 |
|  | 0,483 | 1689 | 1995 | 0,015 | 0,012 | 0,012 | 0,029 | 0,076 | 0,055 | 0,100 | 21-mar-20 | 1,11 |
|  | 0,330 | 1995 | 732 | 0,015 | 0,008 | 0,007 | 0,024 | 0,066 | 0,039 | 0,100 | 30-mar-20 | 0,82 |
| Liguria | 0,643 | 900 | 1933 | 0,037 | 0,010 | 0,040 | 0,030 | 0,070 | 0,100 | 0,062 | 12-mar-20 | 1,52 |
|  | 0,398 | 2126 | 1053 | 0,037 | 0,010 | 0,023 | 0,012 | 0,092 | 0,000 | 0,100 | 21-mar-20 | 0,92 |
| Lombardy | 0,742 | 1679 | 30797 | 0,022 | 0,044 | 0,033 | 0,008 | 0,073 | 0,000 | 0,052 | 27-feb-20 | 1,69 |
|  | 0,308 | 28715 | 5774 | 0,022 | 0,017 | 0,023 | 0,017 | 0,059 | 0,000 | 0,048 | 19-mar-20 | 0,84 |
| Marche | 0,231 | 1906 | 1206 | 0,010 | 0,080 | 0,016 | 0,022 | 0,047 | 0,009 | 0,000 | 05-apr-20 | 0,66 |
|  | 0,133 | 1178 | 311 | 0,010 | 0,007 | 0,011 | 0,000 | 0,057 | 0,002 | 0,068 | 16-apr-20 | 0,42 |
| Molise | 0,217 | 120 | 13 | 0,013 | 0,000 | 0,012 | 0,067 | 0,018 | 0,000 | 0,043 | 02-apr-20 | 0,56 |
| Piedmont | 0,398 | 6527 | 7244 | 0,022 | 0,000 | 0,019 | 0,010 | 0,073 | 0,000 | 0,100 | 23-mar-20 | 1,05 |
|  | 0,363 | 7244 | 4588 | 0,022 | 0,014 | 0,021 | 0,021 | 0,071 | 0,000 | 0,100 | 07-apr-20 | 0,90 |
| Sardinia | 0,296 | 618 | 487 | 0,013 | 0,000 | 0,022 | 0,064 | 0,017 | 0,026 | 0,100 | 24-mar-20 | 0,78 |
|  | 0,216 | 487 | 60 | 0,013 | 0,038 | 0,021 | 0,066 | 0,017 | 0,015 | 0,063 | 02-apr-20 | 0,56 |
| Sicily | 0,402 | 1025 | 952 | 0,015 | 0,000 | 0,016 | 0,048 | 0,055 | 0,034 | 0,100 | 23-mar-20 | 0,93 |
|  | 0,293 | 952 | 271 | 0,015 | 0,000 | 0,009 | 0,017 | 0,068 | 0,012 | 0,100 | 02-apr-20 | 0,75 |
| Trentino | 0,406 | 2305 | 2867 | 0,029 | 0,000 | 0,035 | 0,048 | 0,024 | 0,001 | 0,000 | 17-mar-20 | 1,14 |
|  | 0,291 | 2867 | 2261 | 0,029 | 0,000 | 0,032 | 0,032 | 0,038 | 0,004 | 0,100 | 26-mar-20 | 0,83 |
|  | 0,226 | 2261 | 802 | 0,029 | 0,035 | 0,023 | 0,081 | 0,006 | 0,002 | 0,000 | 07-apr-20 | 0,58 |
|  | 0,201 | 802 | 368 | 0,029 | 0,320 | 0,018 | 0,073 | 0,017 | 0,060 | 0,100 | 24-apr-20 | 0,50 |
| Tuscany | 0,552 | 1675 | 2666 | 0,012 | 0,000 | 0,017 | 0,031 | 0,086 | 0,014 | 0,100 | 16-mar-20 | 1,18 |
|  | 0,353 | 2932 | 1690 | 0,012 | 0,000 | 0,014 | 0,046 | 0,062 | 0,000 | 0,093 | 26-mar-20 | 0,79 |
|  | 0,317 | 1690 | 482 | 0,012 | 0,064 | 0,019 | 0,063 | 0,055 | 0,001 | 0,008 | 12-apr-20 | 0,68 |
| Umbria | 0,134 | 794 | 19 | 0,010 | 0,141 | 0,008 | 0,089 | 0,000 | 0,052 | 0,000 | 26-mar-20 | 0,34 |
| Veneto | 0,807 | 848 | 3538 | 0,031 | 0,000 | 0,018 | 0,062 | 0,052 | 0,000 | 0,100 | 06-mar-20 | 1,75 |
|  | 0,520 | 3538 | 5089 | 0,031 | 0,000 | 0,019 | 0,078 | 0,044 | 0,000 | 0,039 | 16-mar-20 | 1,08 |
|  | 0,336 | 5089 | 1749 | 0,031 | 0,000 | 0,019 | 0,054 | 0,048 | 0,002 | 0,100 | 28-mar-20 | 0,78 |

**Table S4 [Model parameters].** Values of estimated parameters for each region at the end of the identification process. Dates are given corresponding to breakpoints between estimation windows that are automatically detected by the estimation procedure we proposed. Regional net reproduction number are reported in the last column clearly showing the increasing effect of the national lockdown measures taken by the government on March 8[th] 2020. The parameter values used to simulate the exit from the Lock-Down considered in this paper are those shaded in grey for each region corresponding to the estimated parameter in the last window ending on May 3[rd] 2020.

| Source | Model | $\rho$ | | | $\beta$ | | | $\rho \cdot \beta$ (S → I) | | | $\alpha$ (I → Q) | | | $\gamma$ (I → R) | | | $\psi$ (I → H) | | | $\kappa^H$ (Q → H) | | | $\kappa^Q$ (H → Q) | | | $\eta^Q$ (Q → R) | | | $\eta^H$ (H → R) | | | $\zeta$ (H → D) | | | $R_0$ | | |
|---|---|---|---|---|---|---|---|---|---|---|---|---|---|---|---|---|---|---|---|---|---|---|---|---|---|---|---|---|---|---|---|---|---|---|---|---|
| | | nom | min | max | nom | min | max | nom | min | max | nom | min | max | nom | min | max | nom | min | max | nom | min | max | nom | min | max | nom | min | max | nom | min | max | nom | min | max | nom | min | max |
| our model | SIQHRD | variable | 0,122 | 1 | 0,4 | 0,4 | 0,4 | 0,114 | 0,049 | 0,159 | 0,044 | 1E-15 | 0,089 | 0,07 | 0,07 | 0,07 | 0,033 | 8,68E-11 | 0,092 | 0,006 | 1,82E-15 | 0,079 | 0,045 | 8,26E-14 | 0,1 | 0,018 | 0,01 | 0,037 | 0,016 | 1E-19 | 0,32 | 0,02 | 0,006 | 0,029 | 0,78 | 0,195 | 2,272 |
| [S6] | SEPIAHQRD | - | - | - | 0,301 | 0,273 | 0,33 | - | - | - | 0,099 | 0,093 | 0,104 | - | - | - | 0,148 | 0,13986 | 0,156 | - | - | - | - | - | - | 0,07 | 0,063 | 0,073 | 0,07 | 0,063 | 0,073 | 0,041 | 0,037 | 0,045 | 3,6 | 3,49 | 3,84 |
| [S3] | SIDARTHE | - | - | - | - | - | - | - | - | - | - | - | - | - | - | - | 0,017 | - | - | 0,027 | - | - | - | - | - | - | - | - | 0,017 | - | - | 0,01 | - | - | - | - | - |
| [S10] | - | - | - | - | - | - | - | - | - | - | - | - | - | - | - | - | - | - | - | 0,04 | - | - | - | - | - | - | - | - | - | - | - | 0,003 | - | - | - | - | - |
| [S11] | SEIHRD | - | - | - | - | - | - | - | - | - | - | - | - | - | - | - | - | - | - | 0,04 | - | - | - | - | - | 0,072 | - | - | 0,034 | - | - | - | - | - | 2,2 | - | - |
| [S12] | - | - | - | - | - | - | - | - | - | - | 0,099 | - | - | - | - | - | - | - | - | 0,04 | - | - | - | - | - | - | - | - | 0,095 | - | - | - | - | - | 2,4 | - | - |
| [S13] | SEIQRDP | - | - | - | - | - | - | 0,99 | 1 | | - | - | - | 0,068 | 0,09 | | - | - | - | - | - | - | - | - | - | 0 | 0,06 | | - | - | - | - | - | - | - | - | - |
| [S14] | SIR | - | - | - | 0,26 | 0,315 | | - | - | - | - | - | - | - | - | - | - | - | - | - | - | - | - | - | - | - | - | - | - | - | - | - | - | - | - | 2 | 4 |
| [S15] | SIQR | - | - | - | 0,373 | - | - | - | - | - | 0,067 | - | - | - | - | - | - | - | - | - | - | - | - | - | - | - | - | - | - | - | - | - | - | - | - | - | - |
| [S2] | - | - | - | - | - | - | - | - | - | - | - | - | - | - | - | - | - | - | - | - | - | - | - | - | - | - | - | - | - | - | - | - | - | - | 3,28 | 1,4 | 6,49 |
| [S4] | SIRD | - | - | - | - | - | - | 0,143 | 0,348 | | - | - | - | - | - | - | - | - | - | - | - | - | - | - | - | - | - | - | - | - | - | - | - | - | - | - | - |
| [S16] | - | - | - | - | - | - | - | - | - | - | - | - | - | - | - | - | - | - | - | - | - | - | - | - | - | - | - | - | - | - | - | - | - | - | - | 2 | 4 |
| [S17] | SEII | - | - | - | 1,12 | 1,04 | 1,18 | 0,43 | 0,27 | 0,71 | - | - | - | - | - | - | - | - | - | - | - | - | - | - | - | - | - | - | - | - | - | - | - | - | 2,38 | 2,04 | 2,77 |
| Mean, min, max → | | - | - | - | 0,598 | 0,26 | 1,18 | 0,43 | 0,143 | 1 | 0,088 | 0,067 | 0,104 | - | - | - | 0,083 | 0,017 | 0,156 | 0,037 | 0,027 | 0,04 | - | - | - | 0,071 | 0 | 0,073 | 0,054 | 0,017 | 0,095 | 0,018 | 0,003 | 0,045 | 2,772 | 1,4 | 6,49 |
| Source | | | | | | | | | | | | | | | | | | Corrispondence in source | | | | | | | | | | | | | | | | | | | |
| [S6] | | $\lambda$ | | | $\delta_E$ | | | - | | | $\zeta \cdot \eta$ | | | $\gamma_I, \gamma_A$ | | | $(1 - \zeta) \cdot \eta$ | | | - | | | - | | | $\gamma_Q$ | | | $\gamma_H$ | | | $\alpha_H$ | | | $R_0$ | | |
| [S3] | | - | | | - | | | $\alpha, \beta, \gamma, \delta$ | | | $\varepsilon, \theta$ | | | $\lambda, \kappa$ | | | $\mu$ | | | $\nu$ | | | - | | | $\rho, \xi$ | | | $\sigma$ | | | $\tau$ | | | - | | |
| [S10] | | - | | | - | | | - | | | - | | | - | | | - | | | sintomi → ricovero | | | - | | | - | | | - | | | ricovero→decesso | | | - | | |
| [S11] | | - | | | - | | | - | | | $T_{inc}$ | | | $T_{inf}$ | | | - | | | time to hospitaliz. | | | - | | | recov. t. mild cases | | | length hosp. stay | | | - | | | $R_0$ | | |
| [S12] | | - | | | - | | | - | | | incubation period | | | - | | | - | | | sympt. to hospital. | | | - | | | - | | | durat. in hospital | | | - | | | $R_0$ | | |
| [S13] | | - | | | - | | | $\beta$ | | | $\delta$ | | | - | | | - | | | - | | | - | | | $\lambda$ | | | - | | | $\kappa$ (time-variant) | | | - | | |
| [S14] | | - | | | $\beta_0$ | | | - | | | - | | | - | | | - | | | - | | | - | | | - | | | - | | | - | | | $R_0$ | | |
| [S15] | | - | | | $\beta$ | | | - | | | $\eta$ | | | - | | | - | | | - | | | - | | | - | | | - | | | - | | | - | | |
| [S2] | | - | | | - | | | - | | | - | | | - | | | - | | | - | | | - | | | - | | | - | | | - | | | $R_0$ | | |
| [S4] | | - | | | - | | | $\beta$ | | | - | | | - | | | - | | | - | | | - | | | - | | | - | | | - | | | - | | |
| [S16] | | - | | | - | | | - | | | - | | | - | | | - | | | - | | | - | | | - | | | - | | | - | | | $R_0$ | | |
| [S17] | | - | | | $\beta$ pre-lockdown | | | $\beta$ post-lockdown | | | - | | | - | | | - | | | - | | | - | | | - | | | - | | | - | | | $R_0$ | | |

*Additional data used in determining parameter values*

| Data | Value | Source |
|---|---|---|
| % of people that die when hospitalised | 0,014 | [S18] |
| % of people that need to be hospitalized | 0,2 | [S11] |
| % of people that die when infected | 0,02 | [S11] |
| Quarantined infected on the 20/03/20 in Italy | 19185 | [S19] |
| Total positive people on the 20/03/20 in Italy | 37860 | [S19] |

**Table S5 [Comparison with the parameters estimated in other papers in the literature for the national aggregate models].** Comparison between the parameter's values we used in our work and those used in other papers proposing national models for the COVID-19 epidemic in Italy that recently appeared in the literature. Notice that, unfortunately, often it is not possible to pin down a specific parameter in a different model that clearly corresponds to one of ours, and vice-versa. This is because of the different meaning that compartments have in the models and the dynamics of people between the compartments which do not always overlap in an unambiguous manner. When we had to use a time constant, say $\tau$, to determine the value of a parameter, say $k$, that describes the rate of change form compartment $A$ to $B$, we proceeded as follows. We set $k = p / \tau$, where $p \in [0, 1]$ is the percentage of people that has been estimated to actually move from compartment $A$ to $B$.